\documentclass[a4,fproof,numbib]{cta-author}

{}
{}
{}
\usepackage{amsmath}
\usepackage{amsfonts}

\usepackage[caption=false,font=footnotesize]{subfig}
\usepackage{mathtools}
\usepackage{nccmath}
\usepackage{multirow}
\usepackage{graphicx} 
\usepackage{tabularx} 
\usepackage{enumerate}

\begin{document}


\title{Extensive frequency response and inertia analysis under high renewable energy source integration scenarios: application to the European interconnected power system}

\author{\au{Ana~Fern\'{a}ndez-Guillam\'{o}n$^{1}$}, \au{Emilio G\'{o}mez-L\'{a}zaro$^{2}$}, \au{\'{A}ngel~Molina-Garc\'{i}a$^{1\corr}$}}

\address{\add{1}{Automatics, Electrical Eng., and Electronic Tech. Dept., Universidad Politecnica de Cartagena, Cartagena, Spain}
\add{2}{Renewable Energy Research Institute and DIEEAC-EDII-AB, Universidad de Castilla-La Mancha, Albacete, Spain}
\email{angel.molina@upct.es}}

\begin{abstract}
Traditionally, power system's inertia has been estimated according to the rotating masses directly connected to the grid. However, a new generation mix scenario is currently identified, where conventional supply-side is gradually replaced by renewable sources decoupled from the grid by electronic converters (i.e., wind and photovoltaic power plants). Due to the significant penetration of such renewable generation units, the conventional grid inertia is decreasing, subsequently affecting both reliability analysis and grid stability.  As a result, concepts such as `synthetic inertia', `hidden inertia' or `virtual inertia', together with alternative spinning reserves, are currently under discussion to ensure power system  stability and reliability. \textcolor{black}{Under this new framework, an algorithm to estimate the minimum inertia needed to fulfil the ENTSO-E requirements for ROCOF values is proposed and assessed under a relevant variety of imbalanced conditions. The additional active power needed to be within the frequency dynamic range is also estimated and determined. Both inertia and additional active power can come from different sources, such as storage solutions, renewable sources decoupled from the grid including some frequency control strategies, interconnections with other grids, or a combination of them. The power system under consideration includes thermal, hydro-power plants, and renewable generation units, in line with most current and future European supply-side power systems. More than 700 generation mix scenarios are identified and simulated, varying the renewable integration, the power imbalance, and the inertia constant of conventional power plants. In fact, the solutions studied here provide important information to ease the massive integration of renewable resources, without reducing the capacity of the grid in terms of stability and response to contingencies.}
\end{abstract}

\maketitle

\section*{\textcolor{black}{Nomenclature}}\label{sec.nomenclature}

{\color{black}{The following nomenclature has been used:

    \begin{table}[h]
	\renewcommand{\arraystretch}{1.4}
	\resizebox{.95\linewidth}{!}{%
	\begin{tabular}{ll}
	$\Delta f$ & Frequency variation \\
	$\Delta P_{add}$ & Additional power \\
	$\Delta P_{g}$& Variation of the supply-side active power\\
	$\Delta P_{load}$& Variation of the demand-side active power\\
	$\Delta P_{L}$ & Power imbalance\\
	$D_{eq}$ & Equivalent damping factor of loads\\
	$f$ & Frequency\\
	$H$ & Inertia constant\\
	$H_{eq}$ & Equivalent inertia constant\\
	$H_{RES}$ & Virtual inertia constant \\
	$H_{S}$ & Synchronous inertia constant\\
	$S_{B}$ & Rated power of power plant/power system\\
	\hline
	aFR & Automatic frequency restoration\\
	DSO & Distribution system operators\\
	ENTSO-E & European network of transmission system operators for electricity\\
	EU & European Union\\
	FC & Frequency containment\\
	IN & Imbalance netting\\
	mFR & Manual frequency restoration\\
	PFR & Primary frequency reserves \\
    PV & Photovoltaic\\
	RES & Renewable energy sources\\
	ROCOF & Rate of change of frequency\\
	TSO & Transmission system operators\\
	VSWT & Variable speed wind turbines\\
        \end{tabular}
      }
    \end{table}
    
\section{Introduction}\label{sec.introduction}

Power system stability analysis currently relies on synchronous machines with rotational masses connected to the grid. These generation units store kinetic energy, which is automatically extracted in response to a sudden loss of generation, slowing down the machine and reducing the grid frequency~\cite{d15}. However, power systems are gradually changing~\cite{hadjipaschalis09}. Fossil fuel problems, such as resource scarcity, increasing prices and geopolitical risks related to import dependency {\cite{european10,liu12,huber14}}, have encouraged most developed countries to promote Renewable Energy Sources (RES) large-scale integration {\cite{tselepis15}}. In Europe, wind, solar and biomass overtook coal power for the first time in 2017~\cite{sandbag18}.  According the current road-maps, 323 GW and 192 GW of wind and {\color{black}{photovoltaic (PV)}} power plants are expected to be installed in Europe by 2030, covering up to 30\% and 18\% of the {\color{black}{European Union (EU)}} demand respectively {\cite{wind17,iea14}}. Some authors affirm that with current technologies, only 50\% of the overall electricity demand can be given by RES~\cite{zakeri15}. By 2030, the EU will be close to achieving this theoretical limit. {\color{black}{As a result, under this European roadmap, it is desirable to analyse the theoretical renewable threshold by considering the international frequency control requirements for a reliable and secure interconnection frequency response. }}

Among the different RES, PV and wind resources ---especially variable speed wind turbines, VSWT~\cite{OchoaCorrea2017}--- are  considered as the two most promising resources for power generation~\cite{SHAH20151423}. However, their intermittent behaviour makes them difficult to integrate into current power systems. In fact, Transmission and Distribution System Operators (TSO/DSO) deal with not only uncontrollable demand, but also oscillated  generation~\cite{green10,rodriguez14,SHAH20151423,zhang17}. Additionally, these resources do not contribute directly to inertia and power system reserves~\cite{tielens12}, as they are electrically decoupled from the grid through power converters~\cite{spahic16}. It significantly reduces the grid effective inertia~\cite{gautam11}, compromising power system stability and modifying their transient response~\cite{delille12}. Moreover, low system inertia is related to $(i)$ a faster rate of change of frequency (ROCOF) and $(ii)$ larger frequency deviations (larger nadir) within a shorter time frame~\cite{daly15}. Therefore, inertia reduction is considered as a relevant problem to the large integration of RES into power systems~\cite{du18}. Subsequently, \cite{xu18}~considers that there is an acute need to study how inertia reduction impacts system dynamic performances. {\color{black}{Some contributions can be found in the specific literature focused on such inertia reduction impacts. The application of probabilistic and risk-based assessment techniques for operation planning in power systems with high wind power generation has attracted high interest from electrical power industry {\cite{6870479}}.}}

{\color{black}{Different wind turbine controllers with virtual inertia are proposed in the specific literature, mostly evaluated under simplified power system modelling and usually including a few imbalance situations. Fu {\em{et al}} {\cite{7984999}} includes only a sudden load decrease and a  sudden decrease as experimental test. Another examples of such test conditions with a sudden load increase/decrease can be found in {\cite{OchoaCorrea2017,7778835,en11040981,8260147}}.  
Six case studies are considered in {\cite{6519514}}, with hypothetical load step rise events. The loss of the largest power plant of thirty energy schedules were simulated in \cite{fernandez2020frequency} under different wind power integration levels. Wang {\em{et al}} {\cite{7371182}} performs statistical analysis of dynamic frequency stability studies, but considering only the current reference incident (a loss of 3 GW). This reference incident ---also known as worst contingency--- is simulated both in over and under frequency to study the impact of synchronous compensators and battery energy storage systems in order to maintain the stability of the power system adding real or virtual inertia {\cite{8836314}}.
Also the impact of large-scale PV power plants have been also analysed. A review focused on power stability challenges can be found in {\cite{SHAH20151423}}. As an example, Bueno {\em{et al}} {\cite{7456535}} defines six scenarios for stability purposes in transmission systems with large PV power plant integration. }}

{\color{black}{However, it is noted that all these studies focus on providing a specific technique for VSWT/PV power plants to participate in frequency control, evaluating the proposal under a reduced number of scenarios from the supply-side and the imbalance conditions. Even though these approaches can be acceptable to improve the frequency response of power systems, authors propose a more general but detailed and precise frequency analysis in terms of nadir and ROCOF. As a result, this papers focuses on estimating and determining the minimum inertia and additional active power to be provided after an imbalance, following the European Network of Transmission System Operators for Electricity (ENTSO-E) recommendations for nadir and ROCOF values. 
Such virtual inertia and additional power can be provided by a variety of complementary solutions/resources, which can be combined in an optimised power system environment.  A similar but simpler studied was performed in \cite{pddotnuschel2018frequency} for the Australian power system. Primary frequency reserves (PFR) requirements were estimated as a function of the system inertia and the maximum power imbalance; following some predefined frequency constraints such as ROCOF and frequency nadir. 
The power system under analysis includes thermal, hydro-power plants, and different RES integration levels, in line with current and future European generation mix scenarios. Moreover, different inertia constant values for the conventional power plants (i.e., thermal and hydro-power) are considered for the analysis. Preliminary results and a simplified previous analysis was addressed by the authors in {\cite{fernandezguillamon-ICCEP19}}. The contributions of the paper are then summarised as follows:
\begin{itemize}
    \item A relevant variety of scenarios are considered for simulation. In total, 720 different scenarios are analysed, resulting from the combination of 5 RES penetration levels (up to 60\%), 16 power imbalance conditions (up to 40\% in line with ENTSO-E recommendations), 3 thermal inertia constant values, and 3 hydro-power inertia constant values (both inertia constants of conventional power plants are usually considered as constant in most of previous works).
    \item An algorithm to estimate the minimum inertia and the additional power to fulfil the ROCOF and nadir values recommended by ENTSO-E is proposed. These inertia and additional power can come from different sources, such as storage solutions (flywheels, batteries, super-capacitors), participation of RES into frequency control, interconnection with other power systems, increased primary frequency control reserves of conventional units, or a combination of them. 
    \item This study gives extensive results regarding the massive integration of RES and current power system characteristics, maintaining the frequency stability of the grid. Subsequently, it would be then  possible to gradually replace conventional power plants providing the same reliability. The proposed analysis can be extended to other international requirements and rules. 
\end{itemize}}}



The rest of the paper is organised as follows: Section~\ref{sec.inertia} discusses the importance of inertia in power system stability and summarises the recommendations regarding nadir and ROCOF developed by ENTSO-E. The methodology is presented in Section~\ref{sec.methodology}, describing the power system modelling and the different scenarios to be simulated. Results are described in Section~\ref{sec.results}. Finally, Section~\ref{sec.conclusion} presents the conclusions.

\section{Inertia for power system stability. ENTSO-E recommendations} \label{sec.inertia}

\subsection{Preliminaries}
Let us consider turbine-synchronous generators submitted to small variations around their steady-state conditions. The mechanical and electrical power of a generation unit can be then related as follows~\cite{boldea15}:
\begin{equation}\label{eq.gp1}
\Delta\omega_{r}=\dfrac{\Delta P_{m}-\Delta P_{e}}{2\;H\;s+D}  ,
\end{equation}
where $\omega_{r}$ is the rotational pulsation of the generator, $P_{m}$ is the mechanical power, $P_{e}$ represents the electrical power, $H$ is the inertia constant, and $D$ is the damping factor of the loads. $H$ is defined in the specific literature as the time interval (in seconds) during which the generator can supply its rated power from the stored kinetic energy. This time interval usually lies between 2 and 10~s for conventional generators~\cite{tielens16,fernandez17}. Regarding eq.~\eqref{eq.gp1}, certain simplifications are usually considered to apply such expression to power systems with conventional generation: $(i)$ loads are reduced to an aggregated load with an equivalent damping factor $D_{eq}$~\cite{chiodo18}; $(ii)$ the synchronous generators are reduced to an equivalent rotating mass $H_{eq}$:
\begin{equation}
H_{eq}=\dfrac{\displaystyle\sum_{i=1}^{CG} H_{i}\cdot S_{B,i}}{S_{B}} ,
\label{eq.Heq}
\end{equation}
where $H_{i}$ is the inertia constant of the $i$-power plant, $S_{B,i}$ is the rated power of the $i$-power plant, $S_{B}$ is the rated power of the power system, and $CG$ is the total number of conventional generators.

Conventional power plants have been replaced by RES over the last ten years, mainly in response to different policies focused on promoting their integration. These actions have subsequently generated a decline in system inertia~\cite{li17design}. Consequently, larger frequency deviations can be suffered after a supply-side and demand imbalance~\cite{nedd17}; being ROCOF also depending on the available inertia~\cite{ulbig15}. In consequence, TSO and DSO require that RES also contribute to ancillary services~\cite{aho12}, especially wind turbines~\cite{kayikcci09}. Indeed,~\cite{toulabi17} affirms that the wind turbine participation in frequency control is inevitable. By considering these requirements, different alternatives providing additional inertia and frequency control from RES can be found in the recent specific literature.  These solutions are commonly known as `hidden', `synthetic', or `virtual' inertial techniques~\cite{sun10,dreidy17,fernandez19}. The equivalent inertia of power systems can be then divided into two different components: $(i)$~synchronous inertia provided by conventional generators $H_{S}$ and $(ii)$~virtual inertia provided by RES electrically decoupled from the grid, $H_{RES}$, 
\begin{equation}
  H_{eq}=\dfrac{\overbrace{\displaystyle\sum_{i=1}^{CG} H_{i}\cdot S_{B,i}}^{H_{S}} + \overbrace{\displaystyle\sum_{j=1}^{VPP} H_{RES,j} \cdot S_{B,j}}^{H_{RES}}}{S_{B}} ,
  \label{eq.hagg2}
\end{equation} 
where $H_{RES,j}$ is the emulated inertia constant corresponding to the $j$--renewable power plant, and $VPP$ the total number of renewable virtual power plants included in the grid under study. Further information about this equivalent inertia definition can be found in~\cite{morren06phd,tielens17,fernandez19analysis}.

\subsection{ENTSO-E recommendations}\label{sec.entsoe}

ENTSO-E has already focused on the high RES integration--low synchronous inertia problem. In fact, several reports have been published for the EU~\cite{entsoe_europe,entsoe_high,entsoe_si,entsoe_rocof,entsoe_frequency,entsoe_note, entsoe_nc,entsoe_explanatory}. In Europe, frequency control has a hierarchical structure, usually organised in a maximum of five layers (from fast to slow timescales):  $(i)$~frequency containment (FC); $(ii)$~imbalance netting (IN); $(iii, iv)$~frequency restoration (automatic (aFR) and/or manual (mFR)) and $(v)$~replacement (R)~\cite{entsoe_europe}. Table~\ref{tab.frequency_control} presents an overview of each layer. IN reduces the simultaneous activation of aFR from different areas and thus, is not included in the table. 
The increase in RES and loads connected through power electronics is currently a major issue, as several problems are involved,  including  
frequency stability, 
voltage stability, and 
power quality~\cite{entsoe_high}. 
In this paper, the authors focus on frequency stability problems due to the reduction of the power systems' inertia. According to~\cite{entsoe_high}, the minimum inertia of a power system should range  between 2--3~s, which is lower than conventional power plants, as  discussed in Section {\ref{sec.inertia}}. Solutions to avoid a huge decrease of system inertia include~\cite{entsoe_high}: $(i)$ real-time restriction of the maximum penetration of RES connected through power electronics; $(ii)$ inclusion of synchronous compensators or flywheels; and $(iii)$ addition of virtual inertia from power electronic interfaced sources.

\begin{table}[tbp]
  \processtable{European frequency control structure\label{tab.frequency_control}}
  {\begin{tabular*}{20pc}{@{\extracolsep{\fill}}lll@{}}\toprule
     {\bf{Type}}  & {\bf{Activation}} & {\bf{Timescale}} \\
     \midrule
     FC & Automatic activation & 0 -- 30 s\\
     aFR & Automatic activation & 30 s -- 15 min\\
     mFR & Semi-automatic or manual activation & 15 min maximum\\
     R & Semi-automatic or manual activation & 15 min minimum\\
     \botrule
  \end{tabular*}}{}
\end{table}

According to the ENTSO-E recommendations, if the integration of generation units and loads connected to the grid through power electronics is higher than  65\%, the grid may suffer from severe problems in terms of frequency deviation~\cite{entsoe_high}. Current power systems should be robust enough to support a load imbalance of 40\%~\cite{entsoe_si}. Consequently,  virtual inertia is proposed for small/isolated power systems/areas (such as Ireland and Great Britain) with a high integration of non-synchronous elements~\cite{entsoe_si}. In terms of nadir and ROCOF requirements, the main challenges that face these grids after an imbalance are~\cite{entsoe_si}:
\begin{itemize}
    \item {\em{High initial ROCOF}}. The grid inertia value determines the initial ROCOF after an imbalance. To reduce this initial ROCOF when some synchronous generators have been replaced by RES, the electronic interfaced sources (such as PV and VSWT) should deliver  virtual inertia without any delay. 
    \item {\em{Low nadir (minimum frequency)}}. It is important to increase nadir frequency in order to avoid a load shedding program. The frequency control response of conventional power plants can be too slow for this target. However, generation units including power electronics or other storage solutions can be highly effective for this aim, due to power electronic flexibility and control.  
\end{itemize}
By considering both challenges, high initial ROCOF is assumed to be more critical than low nadir, as nadir takes some seconds to reach the minimum frequency value~\cite{entsoe_si} and initial ROCOF is reached practically instantaneously. Indeed, a high ROCOF can endanger the secure system operation of a grid due to~\cite{entsoe_rocof}: 
$(i)$ mechanical limitations of synchronous machines; 
$(ii)$ failure of protection systems; or 
$(iii)$ load shedding programs.
The ROCOF limits defined by ENTSO-E depend on the sliding time-window, considering three different limits~\cite{entsoe_rocof}: 
\begin{itemize}
\item $\pm$ 2 Hz/s for a sliding time-window of 0.5~s 
\item $\pm$ 1.5 Hz/s for a sliding time-window of 1~s 
\item $\pm$ 1.25 Hz/s for a sliding time-window of 2~s 
\end{itemize}
With regard to supply-side generators, they can only be disconnected $(i)$ if any of such ROCOF limits is achieved or $(ii)$ if the frequency deviation is below 47.5~Hz or above 51.5~Hz~\cite{entsoe_rocof}. Nevertheless, under such large frequency deviations, it is difficult to avoid a power system blackout~\cite{entsoe_frequency}. {\color{black}{The dynamic range allowed for frequency deviations is currently $\pm800$~mHz. However, and according to~\cite{entsoe_frequency}, future power systems should handle 2~Hz/s ROCOF and 40\% power imbalance. Under these circumstances, the considered scenarios cover a variety of imbalances (up to 40\%) in line with these recommendations, see Section {\ref{sec.results}}.}}

\section{Methodology}\label{sec.methodology}

\subsection{Power system modelling}\label{sec.cases_study}

With the aim of analysing the influence on the frequency response of high integration of generation units connected to the grid through power electronics, an equivalent power system with a total capacity of 1000~MW is first proposed for simulations. The supply-side mix is in line with most current European power systems, where conventional and renewable units can be considered as an averaged common generation scenario. The system consists of conventional power plants (reheat thermal and hydro-power) and non-dispatchable RES (VSWT and/or PV power plants).  Both thermal and hydro-power plants are modelled by using simplified governor-based models, derived from speed-governing systems widely used and described in~\cite{kundur94}, see Figure~\ref{fig.conventional}. Different inertia constants for thermal and hydro-power plants ($H_T$ and $H_H$ respectively) considered following the review~\cite{fernandez19}. 
\begin{figure}[tbp]
	\centering
	\subfloat[Thermal power plant]{\includegraphics[width=\columnwidth]{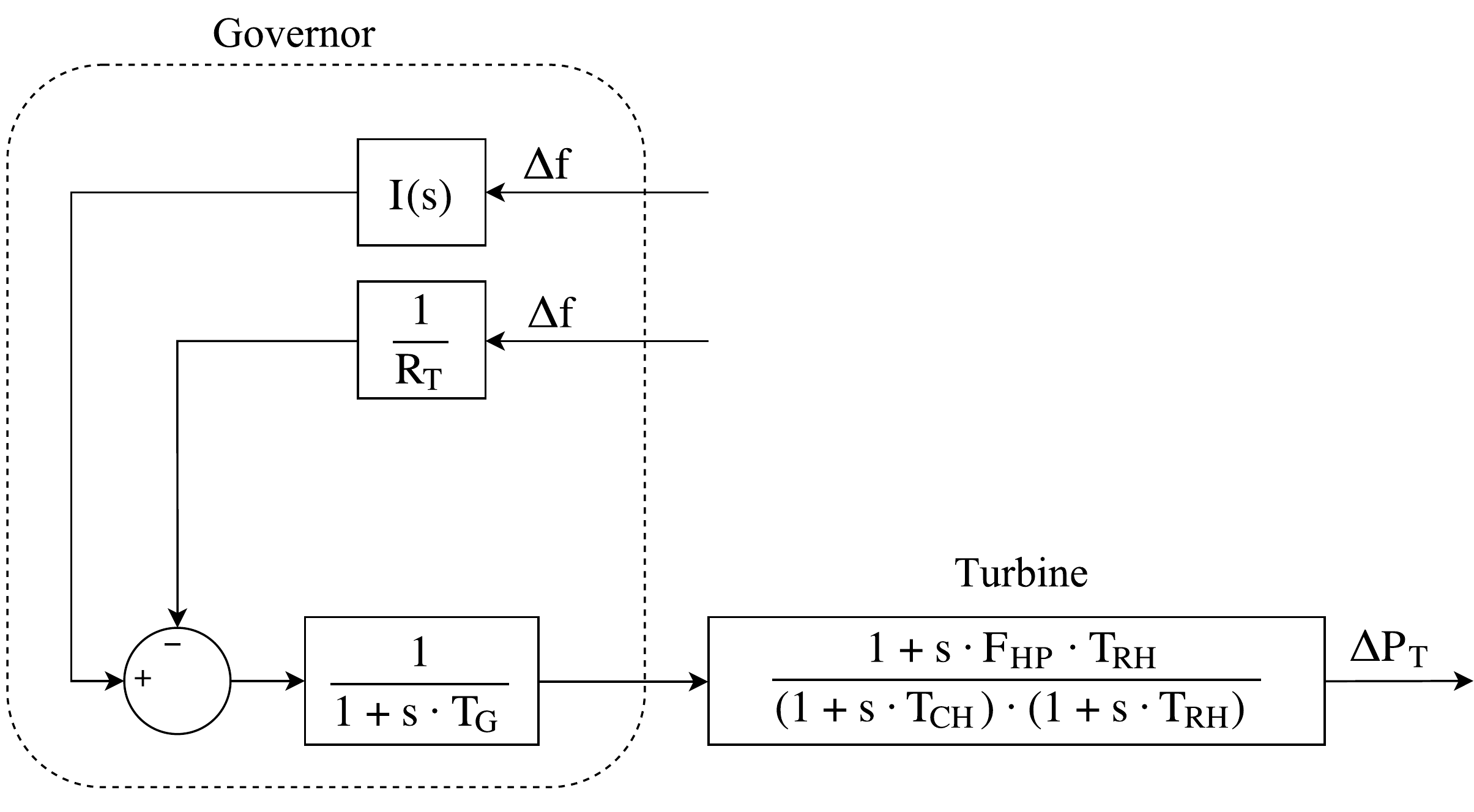}%
		\label{fig.thermal}}
	\hfil
	\subfloat[Hydro-power plant]{\includegraphics[width=\columnwidth]{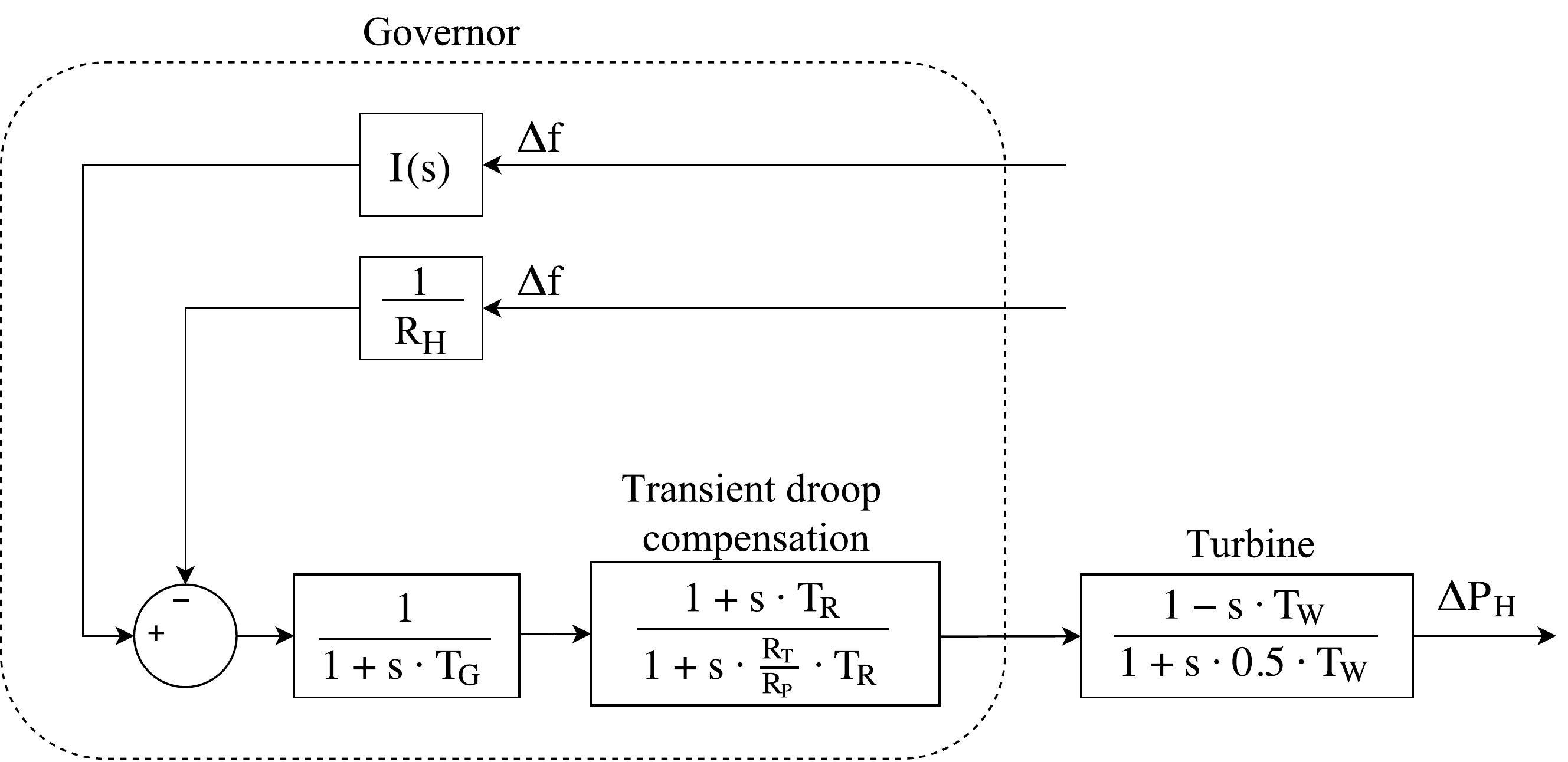}}
		\label{fig.hydro}
	\caption{Conventional power plant models.}
	\label{fig.conventional}
\end{figure}

The power system model for frequency deviation analysis is based on the swing equation, see Section~\ref{sec.inertia}, 
\begin{equation}
\Delta f = \dfrac{1}{2\; H_{eq}\; s + D_{eq}}\cdot (\Delta P_{g}- \Delta P_{load}) ,
\label{eq.swing}
\end{equation}
where $\Delta P_{g}=\Delta P_{RES}+\Delta P_{T}+\Delta P_{H}$ is the active power variation from supply-side and $\Delta P_{load}$ is the demand-side variation. $H_{eq}$ is the equivalent inertia of the power system according to eq. \eqref{eq.hagg2} and $D_{eq}$ is the damping factor, considered as constant $D_{eq}=2$~\%/Hz following the ENTSO-E recommendations~\cite{entsoe_frequency}. {\color{black}{The power imbalance $\Delta P_{L}$ is simulated as a sudden step. Most frequency deviation analysis only consider a reduced number of such sudden imbalances, as was discussed in Section {\ref{sec.introduction}}. To overcome this drawback, 
$\Delta P_{L}$ varies from 2.5\% to 40\% (2.5\% steps) following the ENTSO-E recommendations previously discussed in Section~\ref{sec.entsoe}. }}
\subsection{Virtual inertia and additional power estimation}\label{sec.hres_padd}

Following the case studies explained in Section~\ref{sec.cases_study}, the equivalent synchronous inertia of each simulated scenario is determined by  eq.~\eqref{eq.Heq}. This equivalent inertia lies between 8.71~s (5\% RES, $H_{T}=10$~s and $H_{H}=4.75$~s) and 0.76~s (60\% RES, $H_{T}=2$~s and $H_{H}=1.75$~s). \textcolor{black}{Authors analyse such $H_{eq}$ variations to emphasise  the relevance of such conventional power plant inertia constants. Indeed, small $H$  variations can drastically affect the whole inertia power system $H_{eq}$, and subsequently, the minimum additional virtual inertia to be provided by the other sources.} As will be discussed in Section~\ref{sec.results}, only by considering the synchronous inertia, the ROCOF limits would be violated 
(especially for imbalances $\Delta P_{L}\geq 20$\%) and the nadir frequency would be lower than 49.2~Hz in most cases. Consequently, the virtual additional inertia that should be provided $H_{RES}$ to fulfil the ROCOF limits established by ENTSO-E, as well as the additional power $\Delta P_{add}$ to be within the allowable dynamical variation of frequency, are estimated. Both virtual inertia and additional supply-side power can be provided by a variety of different, complementary solutions/resources, which can be combined in an optimised power system source environment. 

\subsubsection{Estimation of minimum $H_{RES}$}{\label{sec.h_res}}
From the ROCOF limits presented in Section~\ref{sec.entsoe}, the minimum system inertia can be obtained as 
\begin{equation}
    \label{eq.Hmin}
    H_{eq,min}=\dfrac{\Delta P_{L}}{P_{load}}\cdot\dfrac{f_{0}}{ROCOF_{lim}},
\end{equation}
where $\Delta P_{L}$ is the power imbalance, $P_{load}$ is the total load of the power system, $f_{0}$ is the nominal frequency and $ROCOF_{lim}$ is the maximum limit ROCOF allowed~\cite{entsoe_frequency}. Combining eqs.~\eqref{eq.hagg2} and~\eqref{eq.Hmin}, it is possible to obtain the minimum $H_{RES}$ required to fulfil the ROCOF limits:
\begin{equation}\label{eq.hres}
    H_{RES}=\dfrac{H_{eq,min}\cdot S_{B}-H_{T}\cdot S_{B,T}-H_{H}\cdot S_{B,H}}{S_{B,RES}}.
\end{equation}
\textcolor{black}{In consequence, an algorithm to estimate the minimum RES inertia constant $H_{RES}$ is proposed, see Figure~\ref{fig.flow1},}
\begin{figure}[tbp]
    \centering
    \includegraphics[width=0.8\columnwidth]{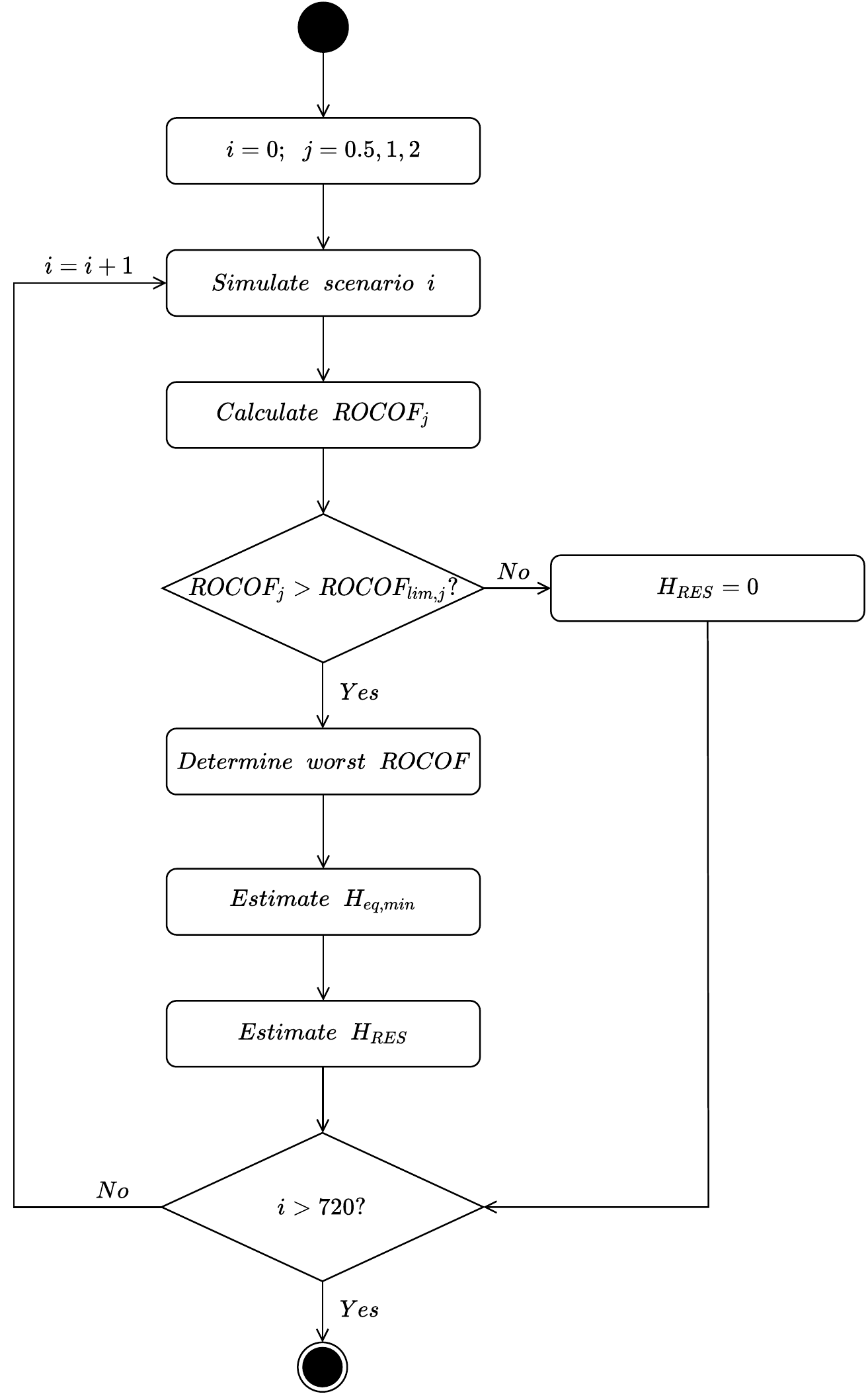}
    \caption{\textcolor{black}{Graphic description to estimate minimum $H_{RES}$ according to the ENTSO-E recommendations.}}
    \label{fig.flow1}
\end{figure}
\begin{enumerate}[i. ]
    \item \textcolor{black}{Simulate the $i$--scenario, depending on the RES percentage, conventional generation units' inertia ($H_{T}$ and $H_{H}$, respectively), and power imbalance $\Delta P_{L}$.
    \item Determine each ROCOF$_{j}$ (being $j=0.5$, $1$ and $2$) with eq.~\eqref{eq.rocof1}--\eqref{eq.rocof3}, where $t_{0}$ is the imbalance time instant (in $s$):
    \begin{equation}
        \label{eq.rocof1}
        ROCOF_{0.5}=\dfrac{\Delta f_{t_{0}+0.5}-\Delta f_{t_{0}}}{0.5},
    \end{equation}
    \begin{equation}
        \label{eq.rocof2}
        ROCOF_{1}=\dfrac{\Delta f_{t_{0}+1}-\Delta f_{t_{0}}}{1},
    \end{equation}
    \begin{equation}
        \label{eq.rocof3}
        ROCOF_{2}=\dfrac{\Delta f_{t_{0}+2}-\Delta f_{t_{0}}}{2}.
    \end{equation}
    \item Check whether any of the previously determined ROCOF$_{j}$ is over their limit: ROCOF$_{0.5}>2$ Hz/s, ROCOF$_{1}>1.5$ Hz/s, and/or ROCOF$_{2}>1.25$ Hz/s. 
    \begin{enumerate}
        \item If no ROCOF$_{j}$ exceeds the corresponding limit, non-additional inertia is required ($H_{RES}=0$), and the following scenario is then simulated ($i=i+1$). 
        \item If one (or more) of the ROCOF$_{j}$ exceed/s its/their limits, the algorithm proceeds to the next step.
    \end{enumerate}
    \item Determine the worst ROCOF$_{j}$ value. This is performed by calculating the relative differences between each ROCOF$_{j}$ exceeding the limits and the corresponding  ROCOF$_{lim,j}$ (in \%). The maximum value of such relative differences is then considered for the next step. 
    \item Estimate the minimum equivalent inertia $H_{eq,min}$ (eq.~\eqref{eq.Hmin}) for the worst ROCOF previously determined, considering ROCOF$_{lim}$ as 2, 1.5 or 1.25~Hz/s depending on ROCOF$_{j}$ (0.5, 1 or 2), respectively.
    \item Calculate the additional RES inertia constant $H_{RES}$ from eq.~\eqref{eq.hres}.
    \item Simulate the following ($i=i+1$) scenario, finishing the algorithm when all scenarios under consideration are simulated ($i>720$).}
\end{enumerate}

\subsubsection{Estimation of minimum $\Delta P_{add}$}\label{sec.p_add} \textcolor{black}{According to ENTSO-E, the maximum dynamic variation of frequency is $f_{0}\pm800$~mHz. Moreover, if $f\leq47.5$~Hz, it is difficult to avoid a power system blackout~\cite{entsoe_frequency}. 
These frequency limits are exceeded in most of the supply-side scenarios currently under analysis. Consequently, some additional power $\Delta P_{add}$ should be provided to be within the dynamic range of such frequency variations. The estimated $\Delta P_{add}$ can be provided by: $(i)$~additional power from RES (by including frequency control strategies), $(ii)$~energy storage solutions, $(iii)$~increasing the PFR of conventional power plants, $(iv)$~other power systems from  interconnections, or $(v)$~a combination of them. Taking into account the power system discussed in Section {\ref{sec.cases_study}}, the corresponding additional active power $\Delta P_{add}$ is then estimated as follows, see Figure \ref{fig.flow2}: 
\begin{figure}[tbp]
    \centering
    \includegraphics[width=0.8\columnwidth]{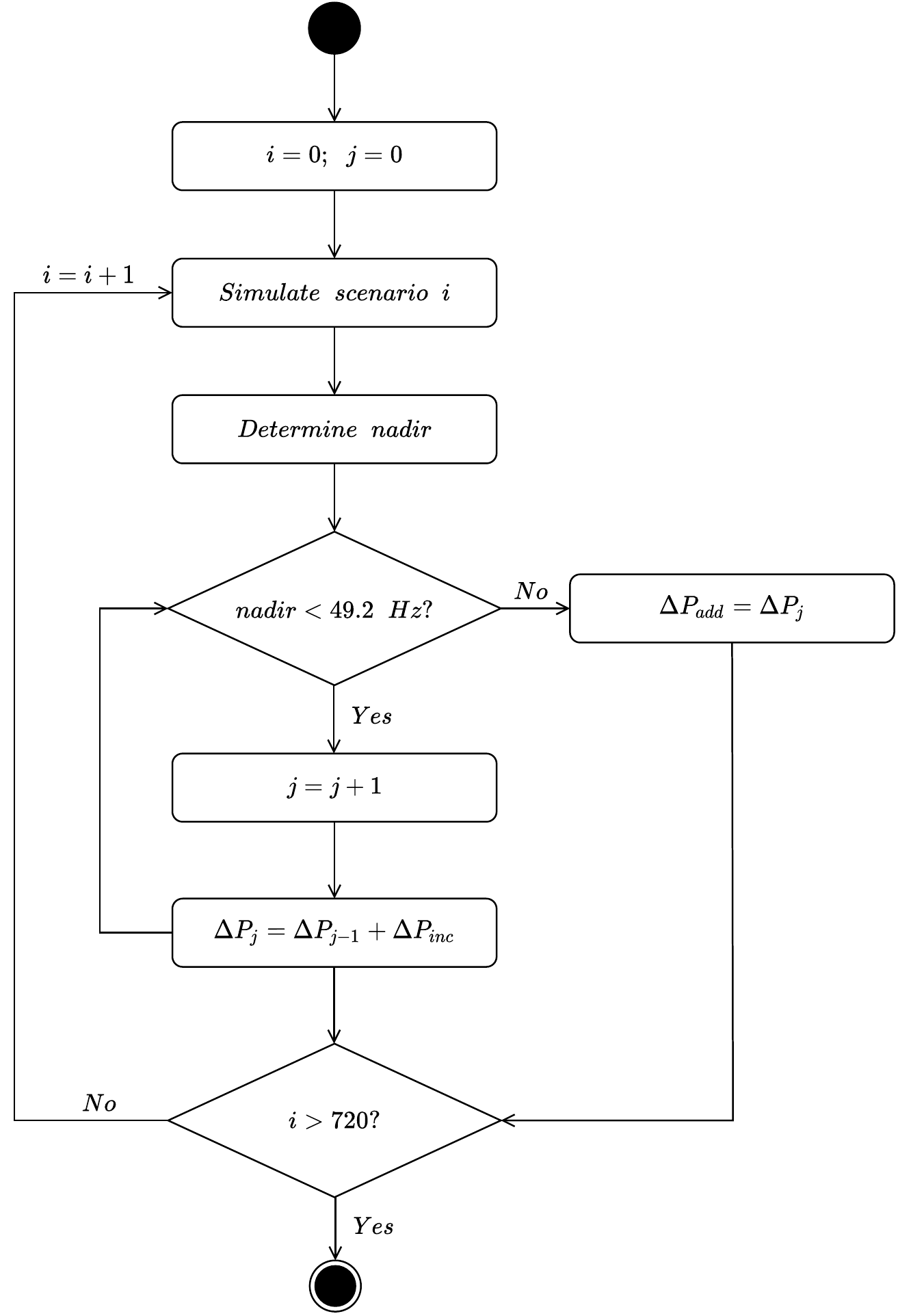}
    \caption{\textcolor{black}{Graphic description to estimate minimum $\Delta P_{add}$ according to the ENTSO-E recommendations.}}
    \label{fig.flow2}
\end{figure}}
\begin{enumerate}[i. ]
    \item \textcolor{black}{Simulate the $i$--scenario, which depends on the RES integration, conventional generation units' inertia ($H_{T}$ and $H_{H}$, respectively) and power imbalance $\Delta P_{L}$.
    \item Determine the nadir (minimum value of the frequency deviation). 
    \item Check if the nadir value is below 49.2 Hz (800 mHz deviation) as recommended by ENTSO-E.
    \begin{enumerate}
        \item If nadir is $f\geq49.2$ Hz, no additional power is needed ($\Delta P_{add}=0$), and a new scenario is simulated ($i=i+1$), going back to the first step of the algorithm. 
        \item If nadir is below 49.2 Hz, the algorithm proceeds to the next step.
    \end{enumerate}
    \item Increase the additional power $\Delta P_{j}$ by a predefined value $\Delta P_{inc}$. In this case, $\Delta P_{inc}=0.01$ pu. This step is repeated until the nadir value is over 49.2 Hz. 
    \item Simulate the following ($i=i+1$) scenario, finishing the algorithm when all scenarios under consideration are simulated ($i>720$).}
\end{enumerate}


\section{Results}\label{sec.results}

\subsection{Case study discussion}
\textcolor{black}{Considering the recommendations of ENTSO-E previously discussed in Section {\ref{sec.entsoe}} as well as current RES integration and roadmaps cited in Section {\ref{sec.cases_study}}, 5 different RES integration levels and 16 different power imbalances are identified for simulation and analysis purposes. {\color{black}{Although the existing studies offer no guidance for the maximum RES limit integration, the percentage to maintain the system frequency response within the permissible bounds is estimated to be around 50\% {\cite{8106748}}, or even higher by other authors with the installation of small, but highly efficient storage devices {\cite{weitemeyer15}}.}} Hydro-power capacity remains constant in all cases. Thermal capacity decreases proportionately as RES capacity increases. The five generation mix cases are summarised in Table~\ref{tab.scenarios}. These generation mix cases are in line with those proposed by the 10-year network development plan for the EU \cite{tyndp2020scenarios}. The power imbalance $\Delta P_{L}$ increases from 2.5\% up to 40\% in 2.5\% steps. }

\begin{table}[tbp]
\renewcommand{\arraystretch}{1.3}
\caption{Generation by source}
\label{tab.scenarios}
\centering
{\begin{tabular*}{20pc}{@{\extracolsep{\fill}}llllll@{}}
\toprule
{\bf{Supply-side resource}} & \multicolumn{5}{c}{\bf{Mix generation (\%)}}  \\
\midrule
Thermal power plant & 80 & 70 & 55 & 40 & 25\\
Hydro-power plant & 15 & 15 & 15 & 15 & 15\\
Renewable power plant & 5 & 15 & 30 & 45 & 60\\
\botrule
\end{tabular*}}{}
\end{table}

{\color{black}{Most of previous contributions discussed in Section {\ref{sec.introduction}} assume a supply-side modelling without any parameter diversity in their equivalent generation units. Following the 
recent review by~\cite{fernandez19}, three different values for the inertia constant of the thermal $H_{T}$ and hydro-power $H_{H}$ plants are considered for simulations. The minimum, mean and maximum value of such inertia constants are then taken into account:  $H_{T}=\left\lbrace 2,6,10\right\rbrace$, $H_{H}=\left\lbrace1.75, 3.25, 4.75\right\rbrace$. Therefore, by combining the different $\Delta P_{L}$ (16), RES integration (5), $H_{T}$ (3) and $H_{H}$ (3), 720 scenarios are  identified and simulated.}} The equivalent synchronous inertia corresponding to the conventional generation units (i.e., thermal and hydro-power) is then determined from eq.~{\eqref{eq.Heq}}, which also depends on the RES integration. All of the simulations are carried out under a Matlab/Simulink environment. 

\subsection{Results}\label{sec.initial_results}

\begin{figure}[htbp]
	\centering
	\subfloat[Nadir frequency]{\includegraphics[height=0.2\textheight]{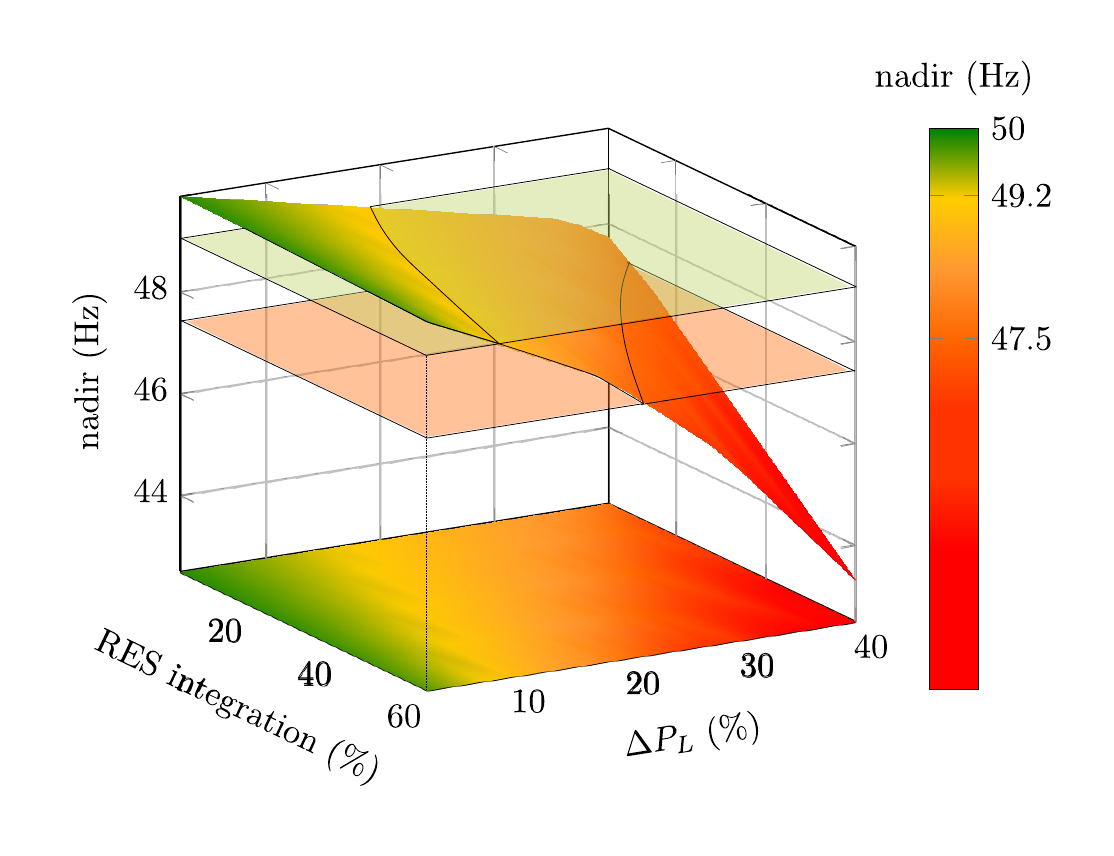}}\\
	\subfloat[ROCOF$_{0.5}$ overview]{\includegraphics[height=0.21\textheight]{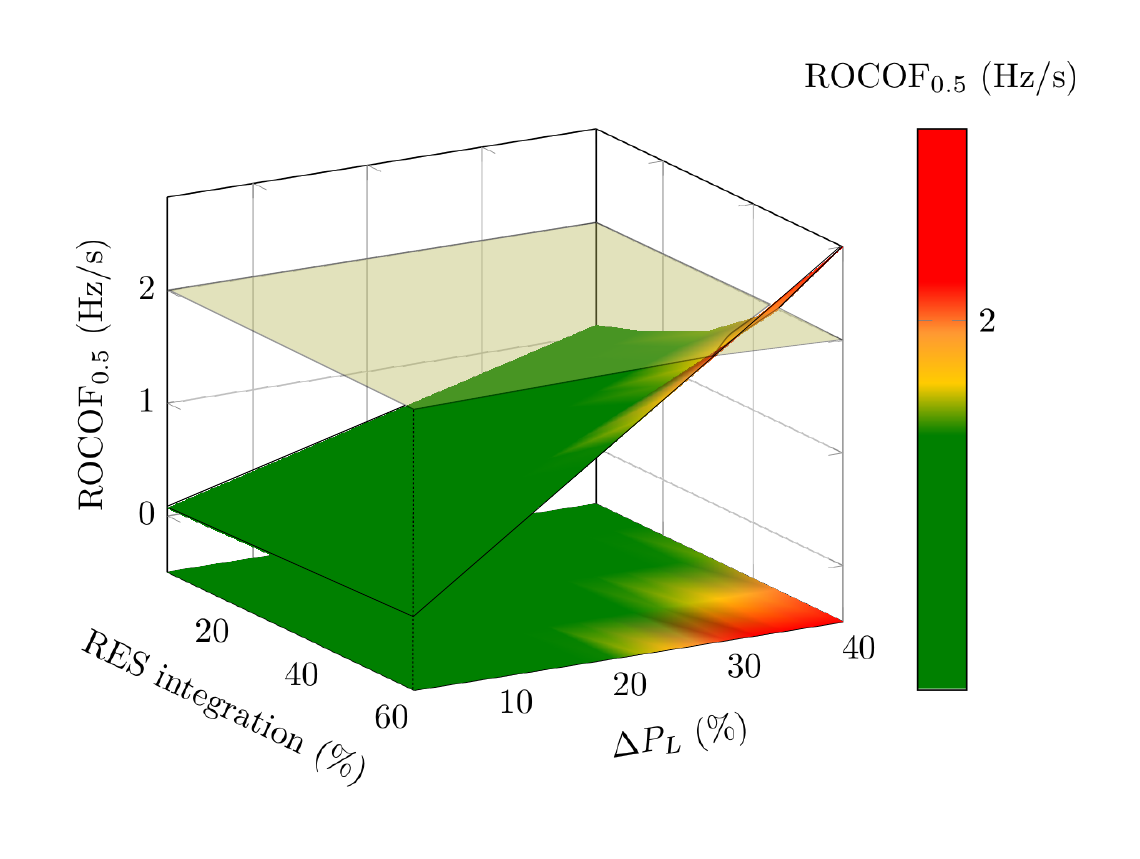}}\\
	\subfloat[ROCOF$_{1}$ overview]{\includegraphics[height=0.21\textheight]{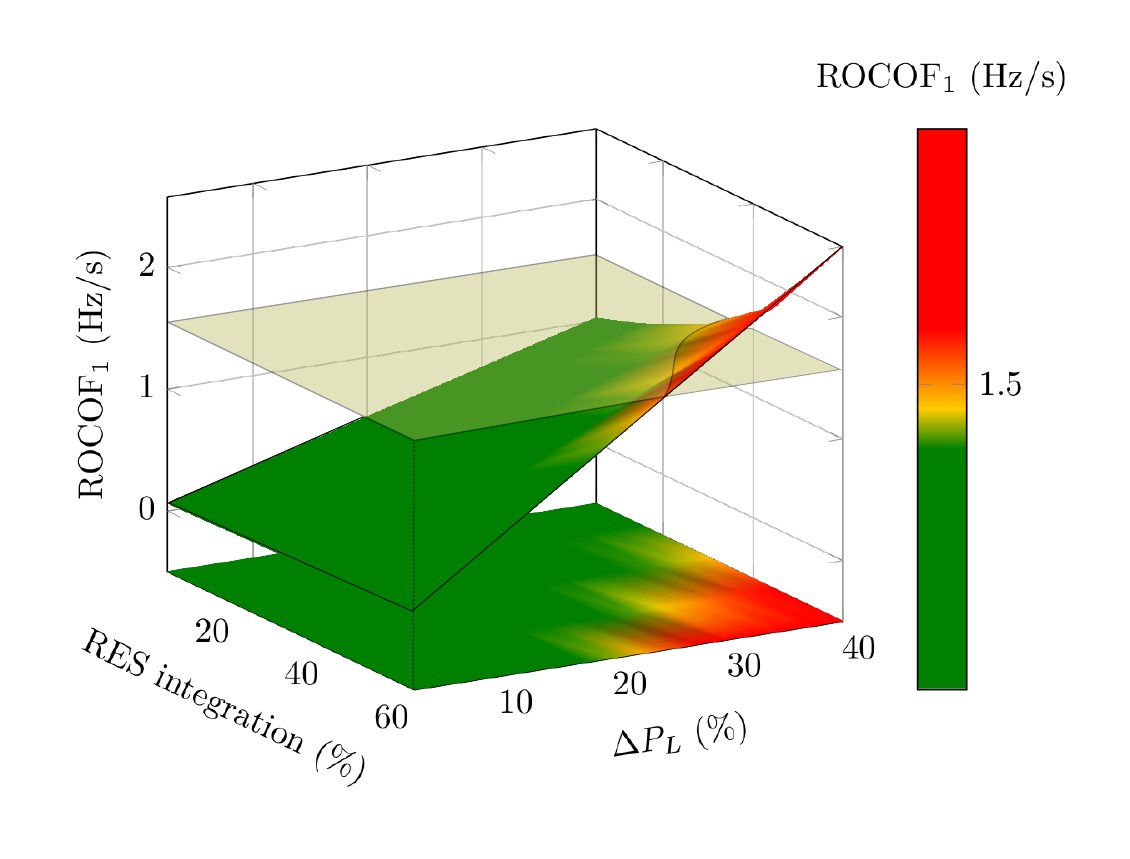}}\\
	\subfloat[ROCOF$_{2}$ overview]{\includegraphics[height=0.21\textheight]{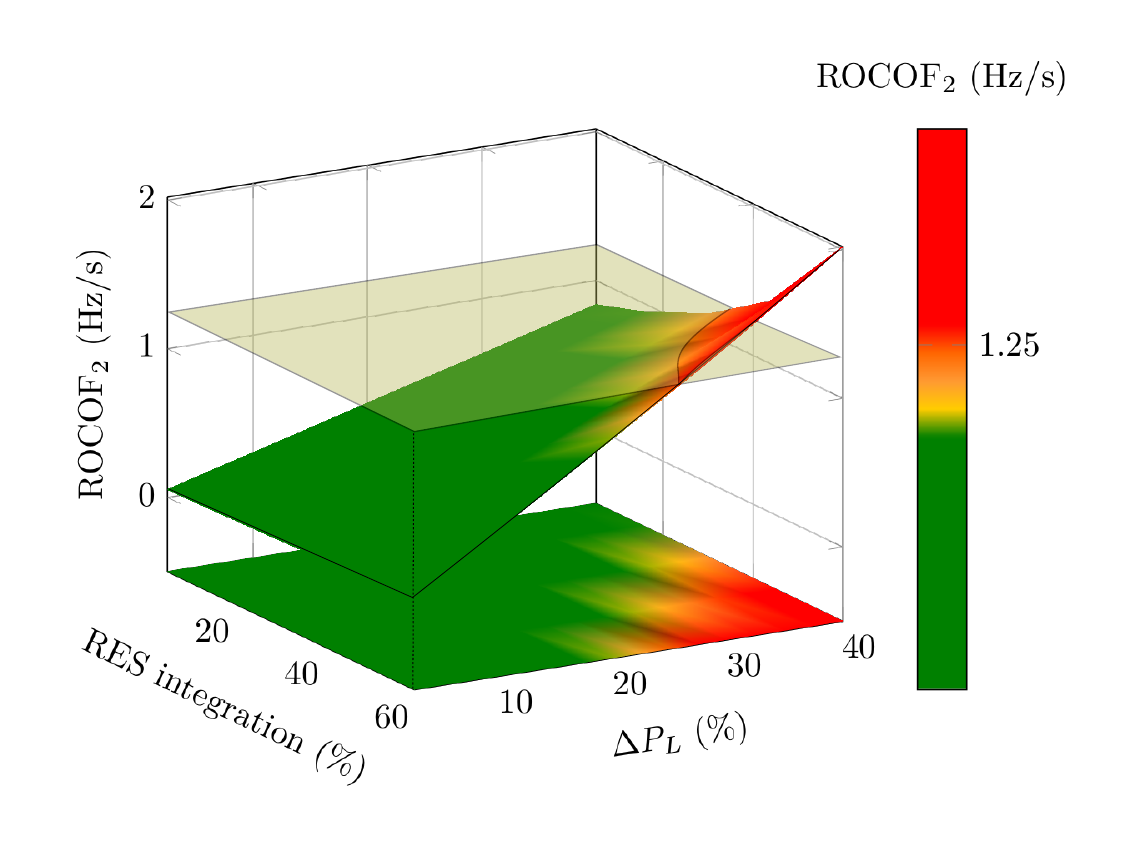}}%
	\caption{\textcolor{black}{Overview of nadir frequency and ROCOFs when $H_{T}=10$~s and $H_{H}=4.75$~s}}
	\label{fig.case1}
\end{figure}

\begin{figure}[htbp]
	\centering
	\subfloat[Nadir frequency]{\includegraphics[height=0.21\textheight]{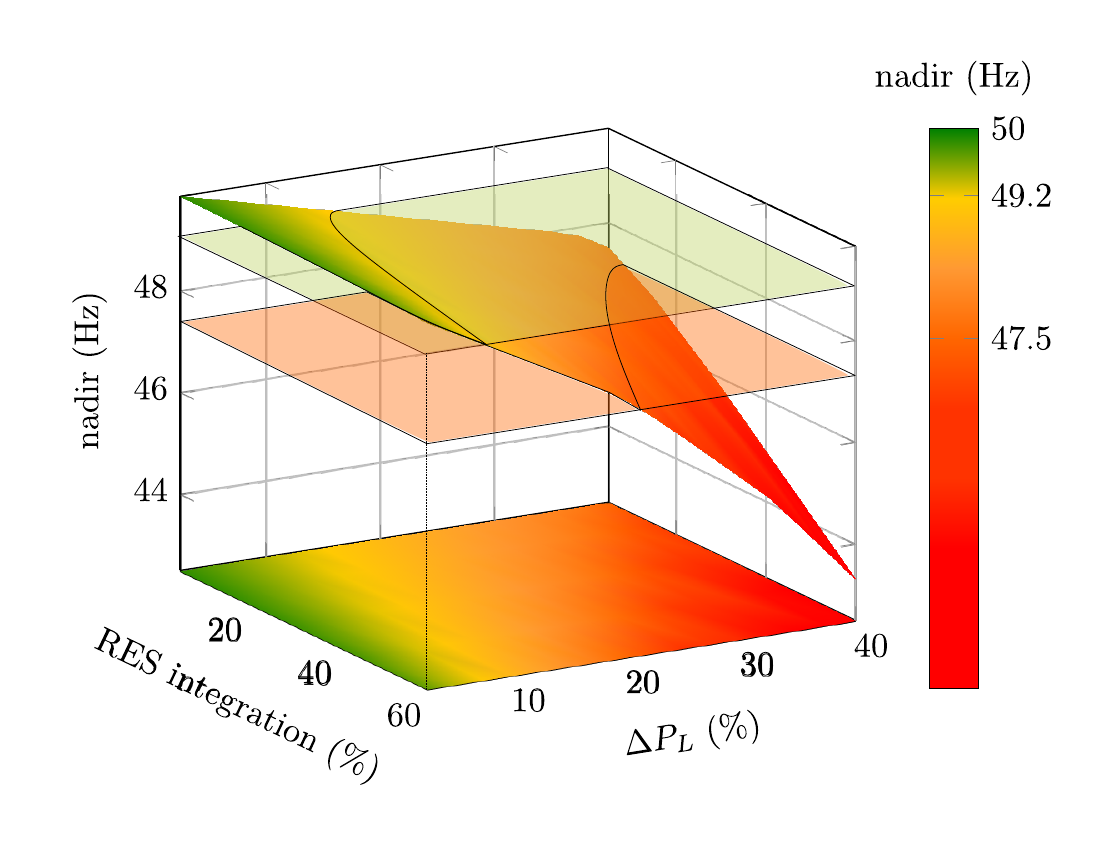}}\\
	\subfloat[ROCOF$_{0.5}$ overview]{\includegraphics[height=0.21\textheight]{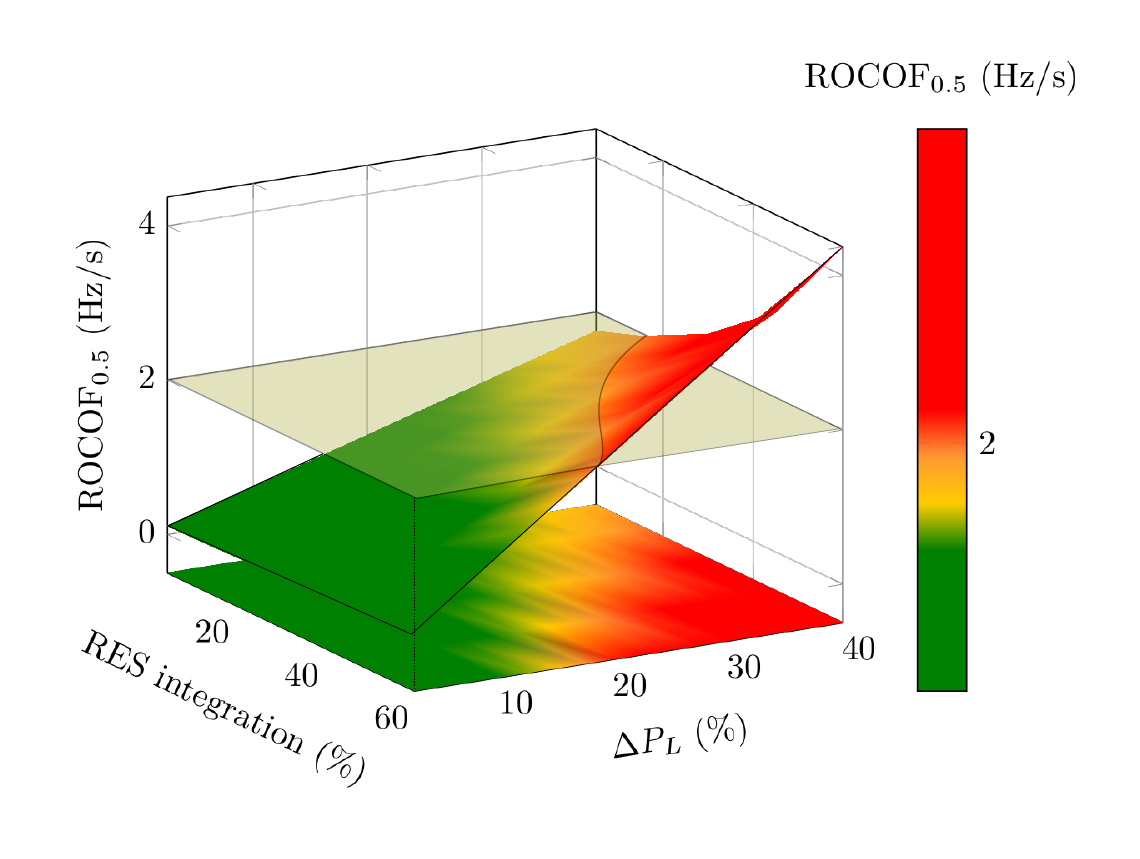}}\\
	\subfloat[ROCOF$_{1}$ overview]{\includegraphics[height=0.21\textheight]{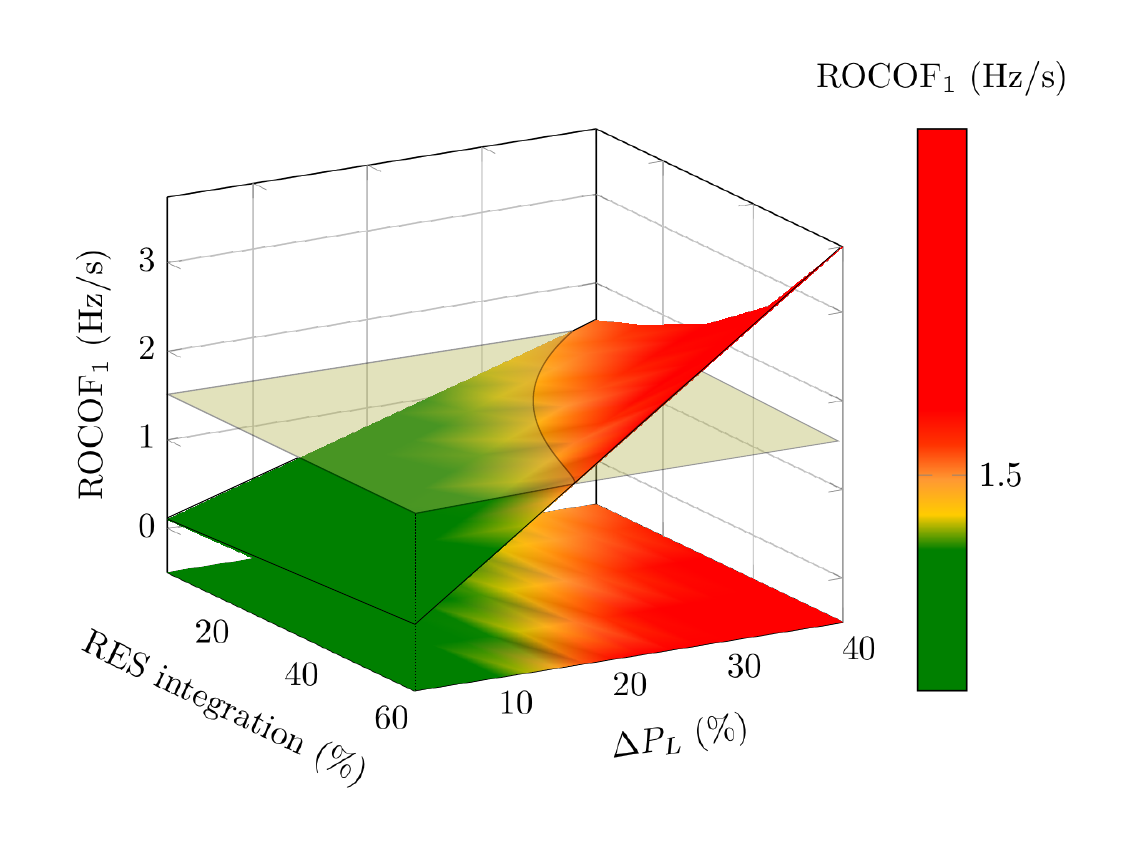}}\\
	\subfloat[ROCOF$_{2}$ overview]{\includegraphics[height=0.21\textheight]{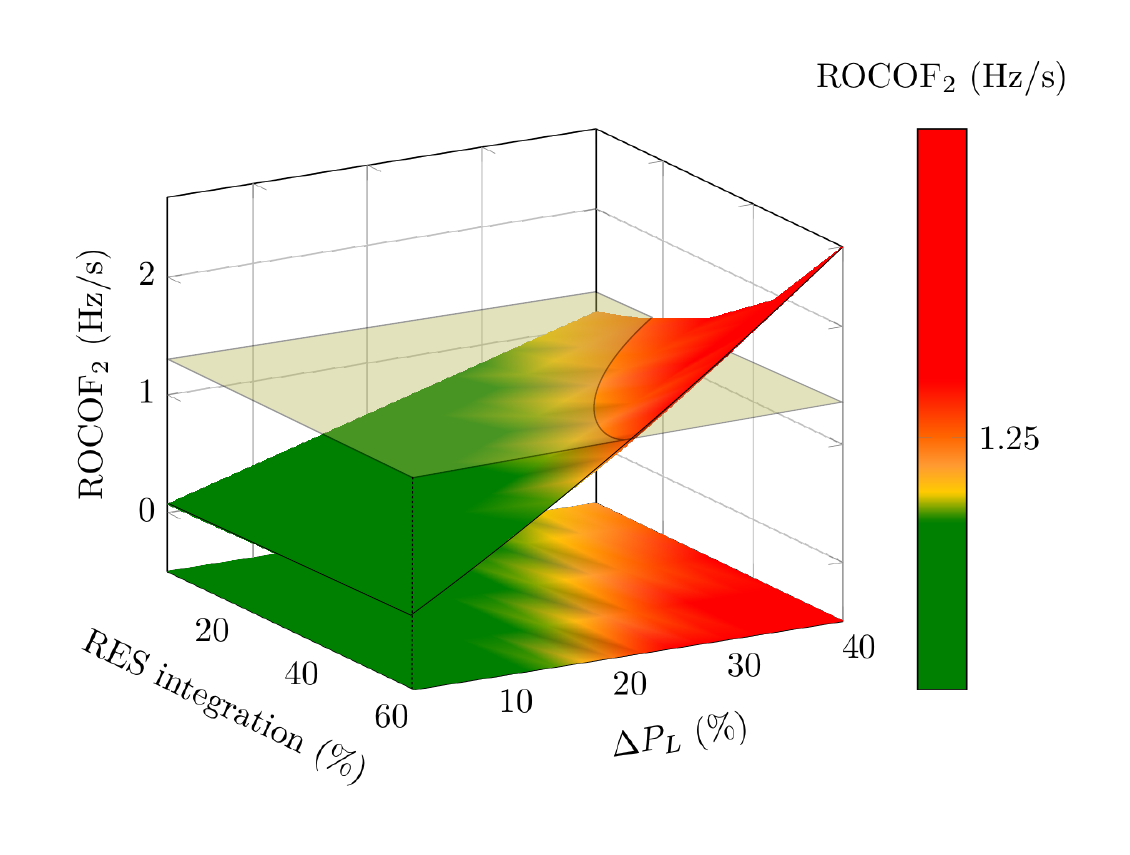}}%
	\caption{\textcolor{black}{Overview of nadir frequency and ROCOFs when $H_{T}=6$~s and $H_{H}=3.25$~s}}
	\label{fig.case2}
\end{figure}

\begin{figure}[htbp]
	\centering
	\subfloat[Nadir frequency]{\includegraphics[height=0.21\textheight]{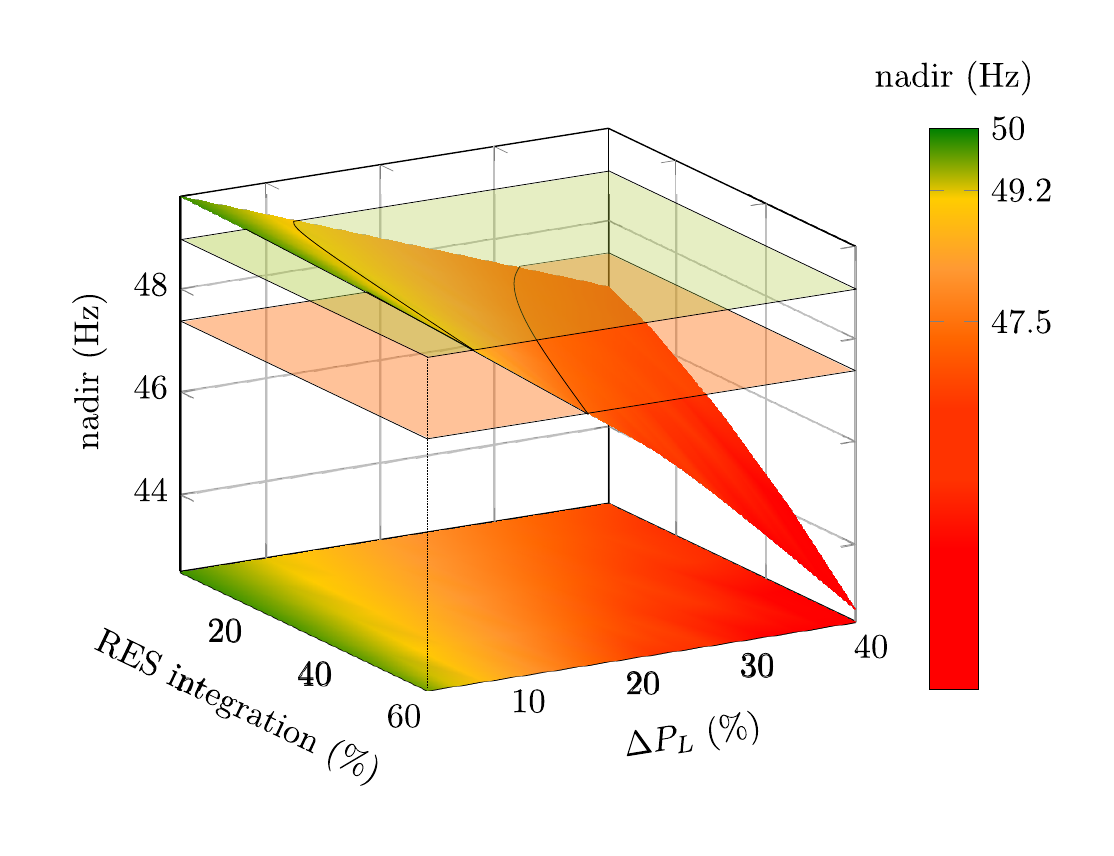}}\\
	\subfloat[ROCOF$_{0.5}$ overview]{\includegraphics[height=0.21\textheight]{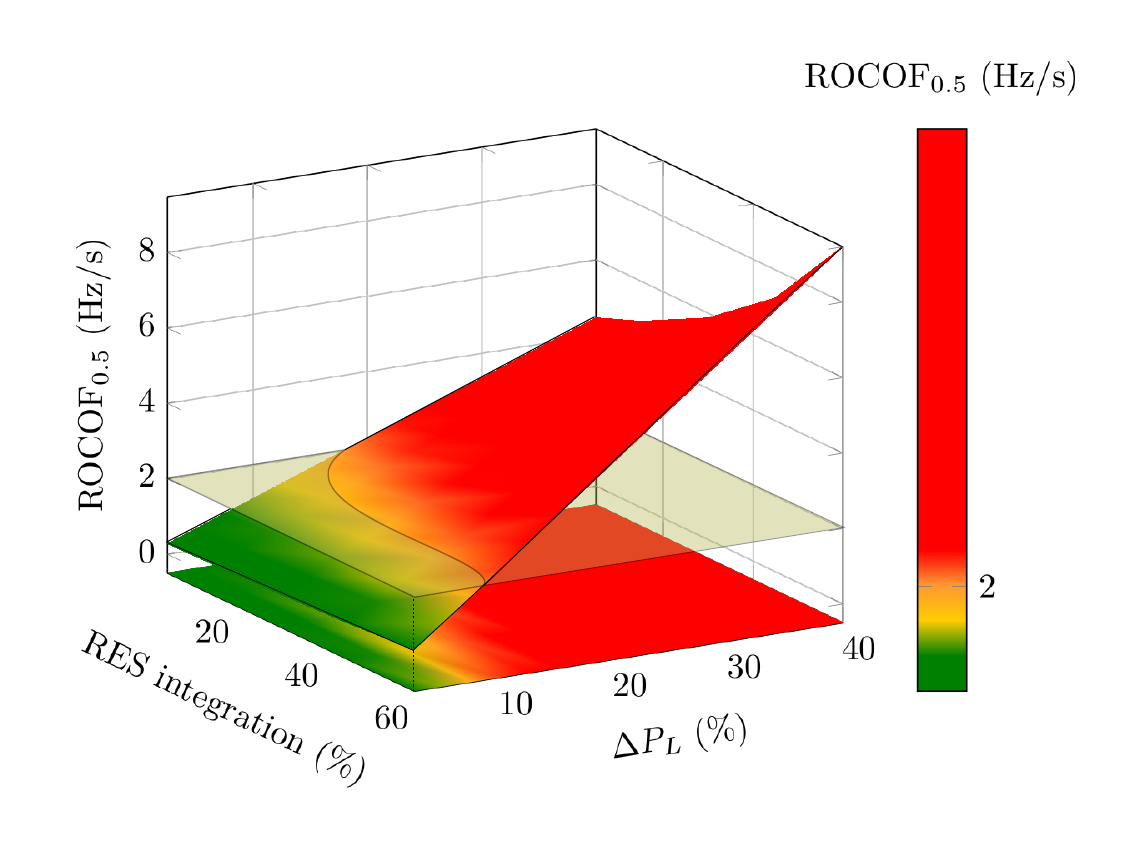}}\\
	\subfloat[ROCOF$_{1}$ overview]{\includegraphics[height=0.21\textheight]{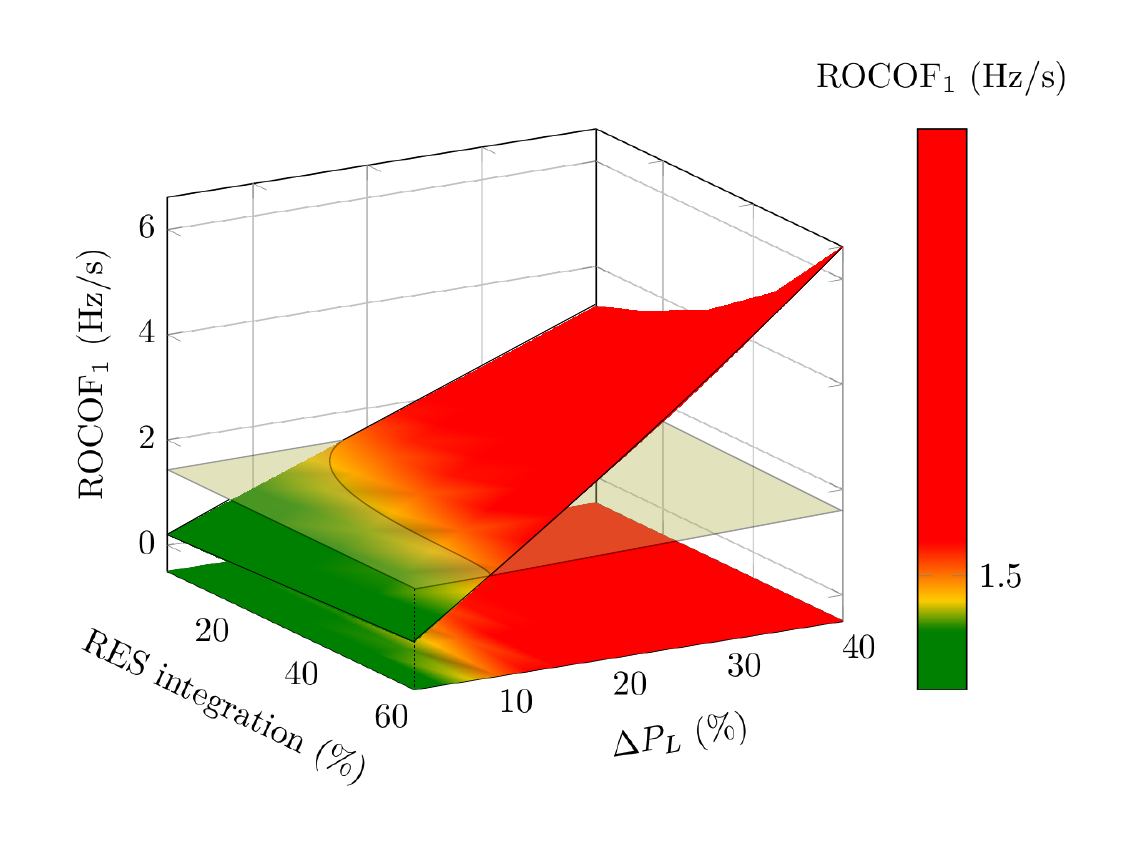}}\\
	\subfloat[ROCOF$_{2}$ overview]{\includegraphics[height=0.21\textheight]{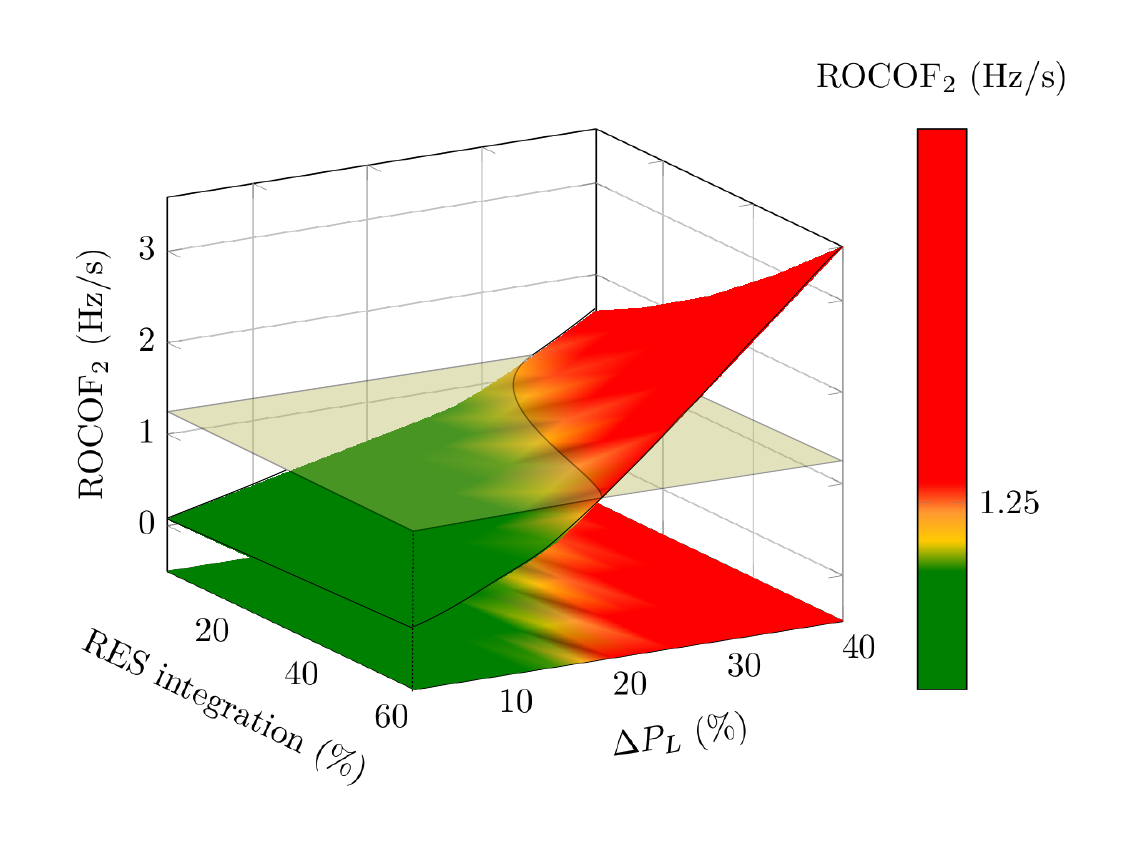}}%
	\caption{\textcolor{black}{Overview of nadir frequency and ROCOFs when $H_{T}=6$~s and $H_{H}=3.25$~s}}
	\label{fig.case3}
\end{figure}

\textcolor{black}{Figures~\ref{fig.case1}--\ref{fig.case3} depict 240 cases out of the 720 under consideration. In each figure, 80 scenarios are summarised (combination of the 16 power imbalances and the 5 generation mix). Both the nadir frequency and the three ROCOF values for three different combinations of $H_{T}$ and $H_{H}$ are depicted, being significant the diversity of results under a variety of such parameters, commonly assumed as constant in most of previous contributions as was discussed in Section {\ref{sec.introduction}}. The equivalent synchronous inertia of the system for each figure is detailed in Table \ref{tab.H_eq}, depending on the RES integration.}

\begin{table}[tbp]
  \processtable{Synchronous equivalent inertia ($H_{eq}$). Example of results\label{tab.H_eq}}
  {\begin{tabular*}{20pc}{@{\extracolsep{\fill}}cccccc@{}}\toprule
     {\bf{}}  &  \multicolumn{5}{c}{\bf{RES Integration (\%)}} \\
     {\bf{}}  & {\bf{5}} &  {\bf{15}} & {\bf{30}}  & {\bf{45}}  & {\bf{60}} \\
     \midrule
     $H_T=10$ & {\multirow{2}{*}{8.71}} & {\multirow{2}{*}{7.71}} & {\multirow{2}{*}{6.21}} & {\multirow{2}{*}{4.71}} & {\multirow{2}{*}{3.21}} \\
     $H_{H}=4.75$ &  &  & &  & \\
          \midrule
      $H_T=6$ & {\multirow{2}{*}{5.29}} & {\multirow{2}{*}{4.69}} & {\multirow{2}{*}{3.79}} & {\multirow{2}{*}{2.89}} & {\multirow{2}{*}{1.99}} \\
     $H_{H}=3.25$ &  &  & &  & \\
          \midrule
      $H_T=2$ & {\multirow{2}{*}{1.86}} & {\multirow{2}{*}{1.66}} & {\multirow{2}{*}{1.36}} & {\multirow{2}{*}{1.06}} & {\multirow{2}{*}{0.76}} \\
     $H_{H}=1.75$ &  &  & &  & \\
     \botrule
  \end{tabular*}}{}
\end{table}

\textcolor{black}{With regard to nadir frequency, values over $f\geq49.2$~Hz are dynamically accepted by ENTSO-E. As a consequence, they are coloured in green. If the nadir decreases under $f<49.2$~Hz (but is still over $f\geq47.5$~Hz), it is coloured in yellow and orange, as they are outside the dynamic range allowed by ENTSO-E. When nadir is under $f<47.5$~Hz, a possible blackout may occur in the grid, subsequently being coloured in red. Two different planes are included in Figures \ref{fig.case1}--\ref{fig.case3} determining such limits. As can be seen, most of the scenarios under consideration are outside the dynamically accepted range, especially if both the RES integration and the power imbalance are high. Even though the equivalent synchronous inertia is significantly reduced depending on the $H_{T}$ and $H_{H}$ values (for instance, $H_{eq}=8.71$ s, $H_{eq}=5.29$ s or $H_{eq}=1.86$ s for 5\% RES integration, refer to Table \ref{tab.H_eq}), the nadir frequency does not decrease drastically for the same generation mix. In fact, if comparing the 5\% RES integration energy mix, and the imbalance $\Delta P_{L}=2.5$\%, the nadir frequencies are 49.88, 49.86 and 49.80 Hz, from the highest to the lowest combination values of $H_{T}$ and $H_{H}$. Similar results are obtained for the worst scenario (i.e., 60\% RES integration energy mix and the imbalance $\Delta P_{L}=40$\%), where the nadir frequencies are 43.33, 43.33 and 42.75 Hz, respectively.}

\textcolor{black}{By considering the three ROCOF values, the limits established by ENTSO-E are ROCOF$_{0.5}\leq2$ Hz/s, ROCOF$_{1}\leq1.5$ Hz/s and ROCOF$_{2}\leq1.25$ Hz/s for the 0.5, 1 and 2 s time-window, respectively. ROCOF values  
are coloured in accordance to their limits,  
in line with the nadir frequency.
Most scenarios under consideration are outside the limits determined, especially if both the RES integration and the power imbalance $\Delta P_{L}$ are high values. In contrast to nadir values, ROCOFs are really dependent on the synchronous equivalent inertia. Indeed, the lower the $H_{eq}$, the higher the ROCOFs values (compare, for instance, the ROCOFs of Figure~\ref{fig.case1} to those values of Figure \ref{fig.case3}).}


\subsection{$H_{RES}$ and $\Delta P_{add}$ estimation}\label{sec.including_hres_padd}

\begin{figure}[tbp]
	\centering
	\subfloat[$\Delta P_{add}$ to fulfil nadir frequency limit]{\includegraphics[height=0.21\textheight]{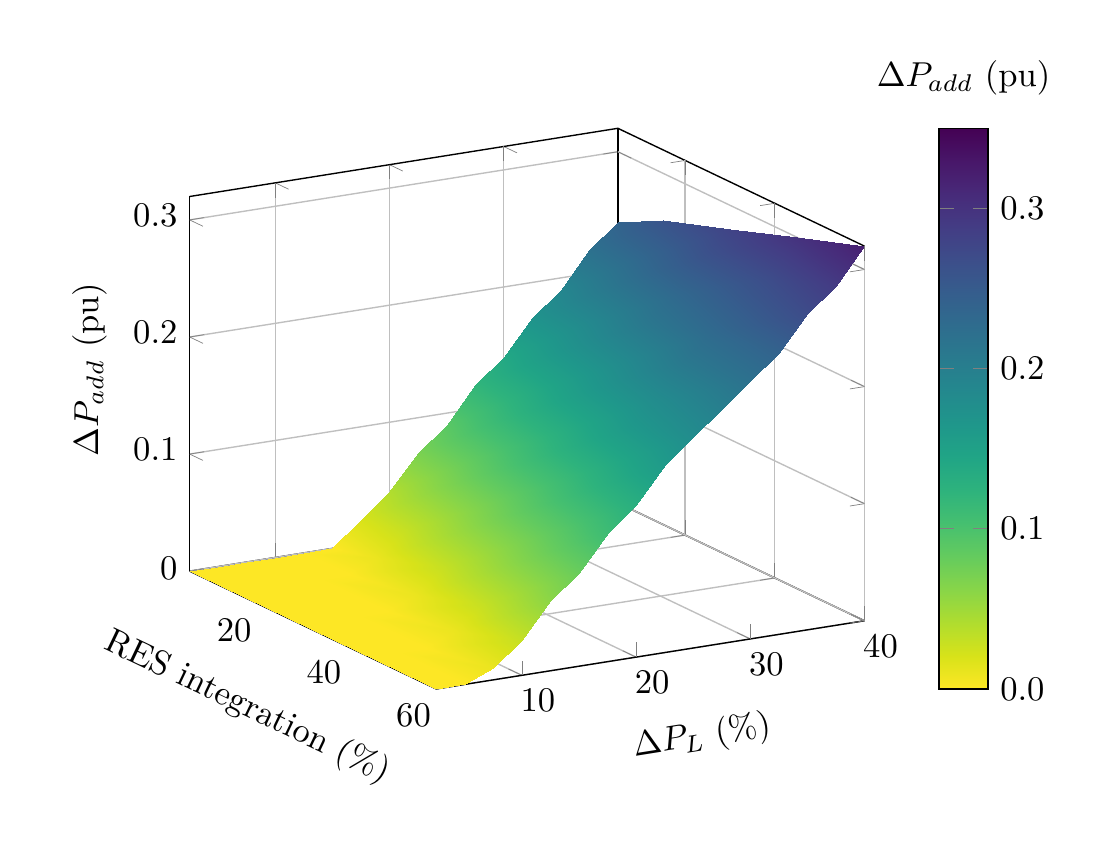}}%
	\hfil
	\subfloat[$H_{RES}$ to fulfil ROCOFs limits]{\includegraphics[height=0.21\textheight]{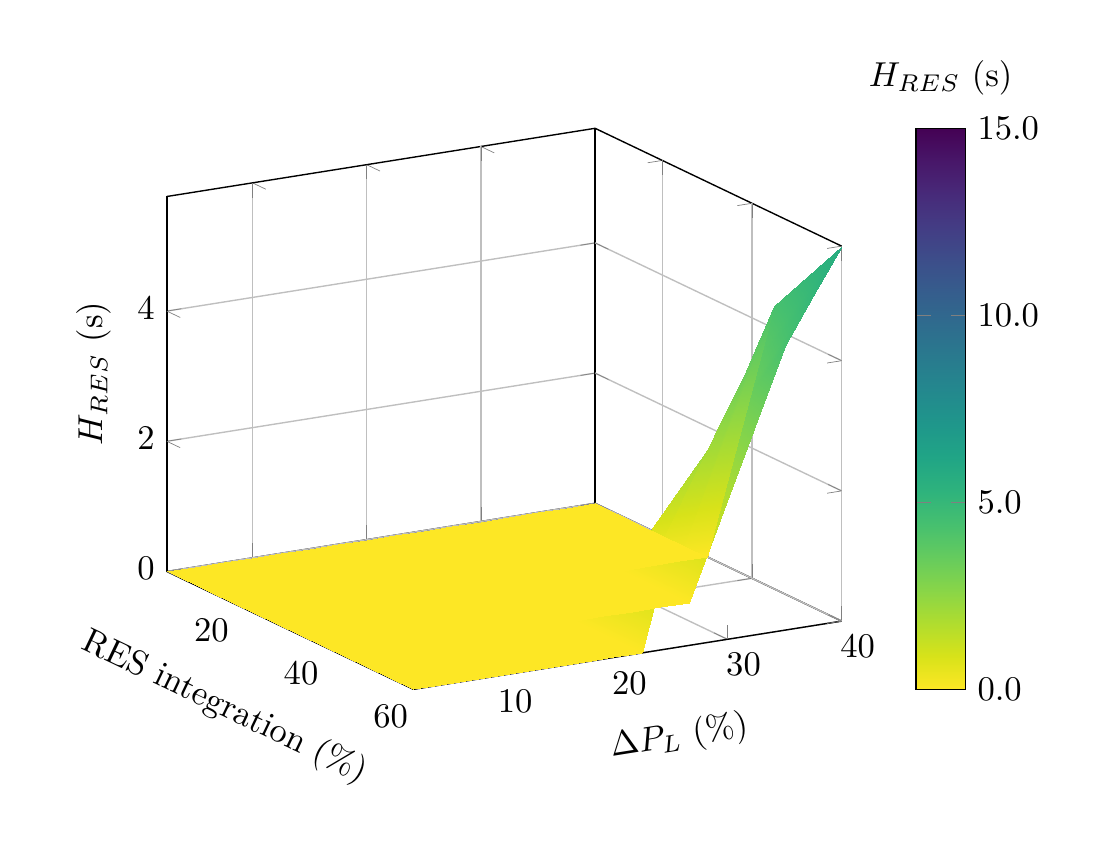}}
	\caption{\textcolor{black}{$H_{RES}$ and $\Delta P_{add}$ estimation ($H_{T}=10$~s, $H_{H}=4.75$~s).
}}
	\label{fig.case1_mod}
\end{figure}

\begin{figure}[tbp]
	\centering
	\subfloat[$\Delta P_{add}$ to fulfil nadir frequency limit]{\includegraphics[height=0.21\textheight]{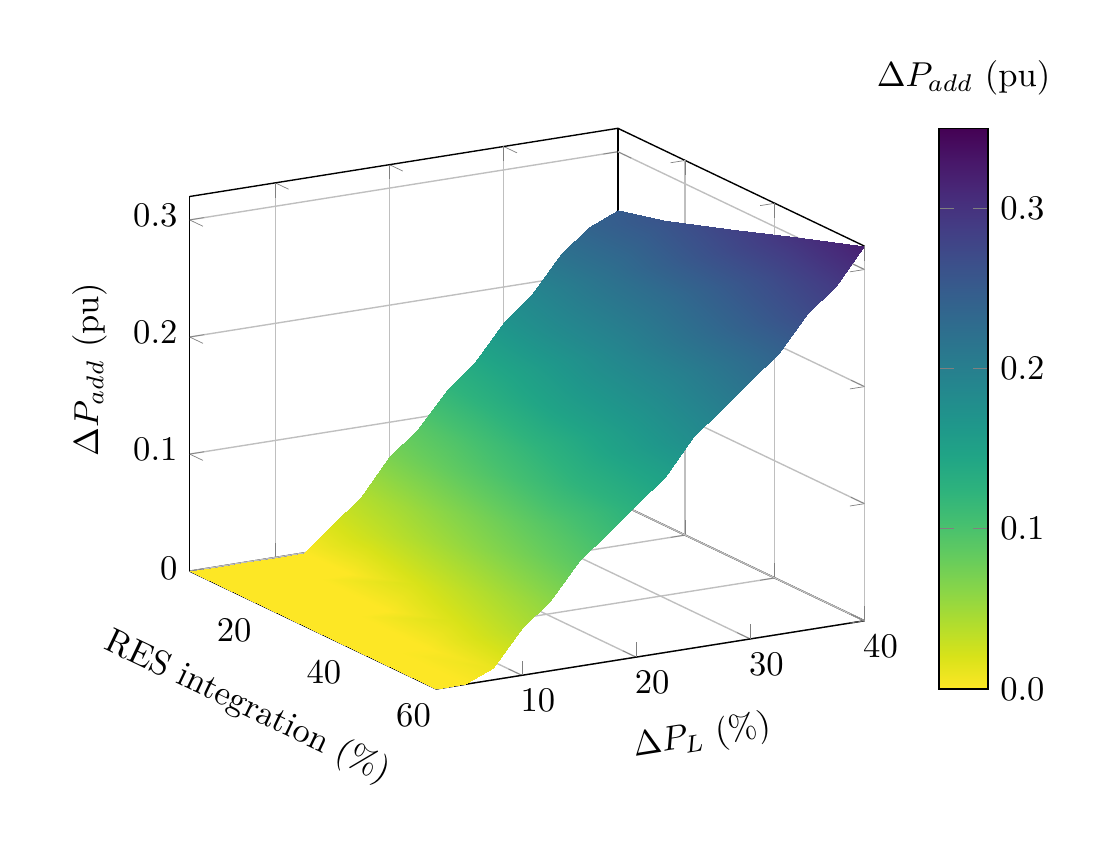}}%
	\hfil
	\subfloat[$H_{RES}$ to fulfil ROCOFs limits]{\includegraphics[height=0.21\textheight]{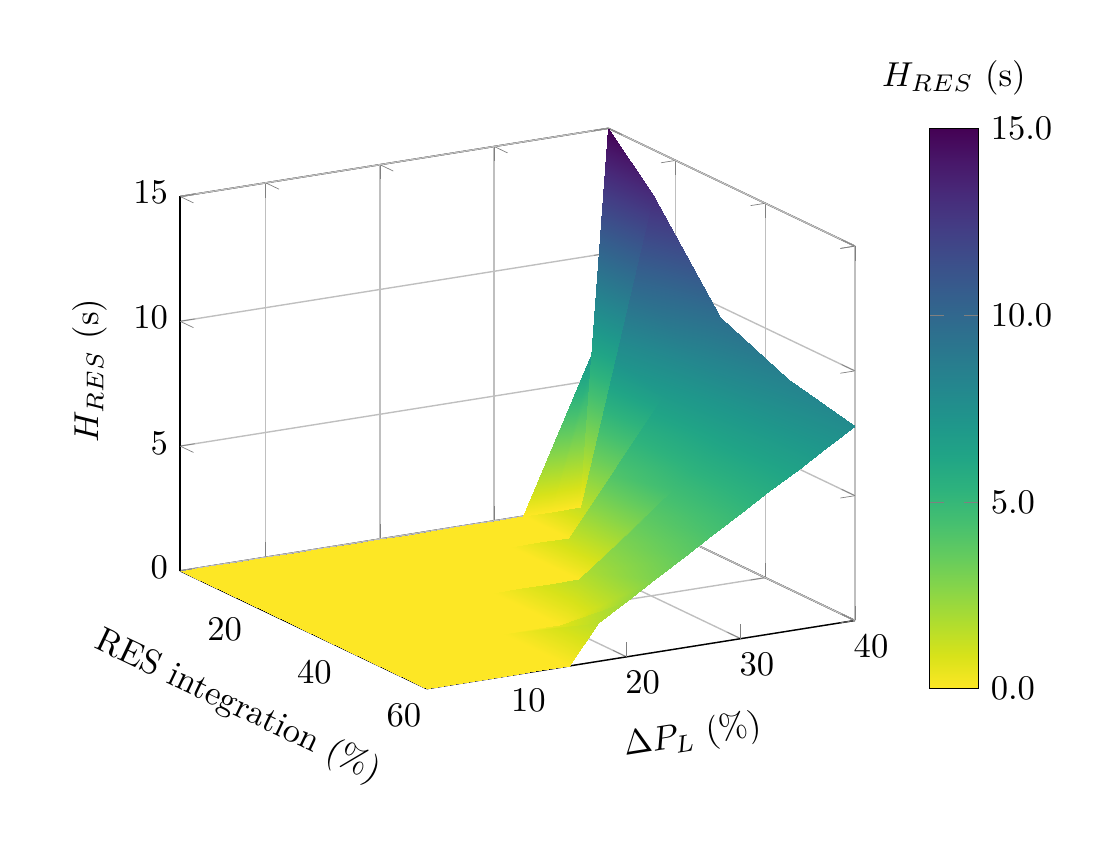}}
	\caption{\textcolor{black}{$H_{RES}$ and $\Delta P_{add}$ estimation ($H_{T}=6$~s, $H_{H}=3.25$~s).}}
	\label{fig.case2_mod}
\end{figure}

\textcolor{black}{Following the methodology described in Section \ref{sec.hres_padd}, the virtual inertia $H_{RES}$ to fulfil the ROCOF limits, and the additional active power $\Delta P_{add}$ needed to be within the dynamic range of frequency variations are  estimated. The results shown in Figures~\ref{fig.case1_mod}--\ref{fig.case3_mod} correspond to the 240 cases presented in Figures~\ref{fig.case1}--\ref{fig.case3}.}

\begin{figure}[tbp]
	\centering
	\subfloat[$\Delta P_{add}$ to fulfil nadir frequency limit]{\includegraphics[height=0.21\textheight]{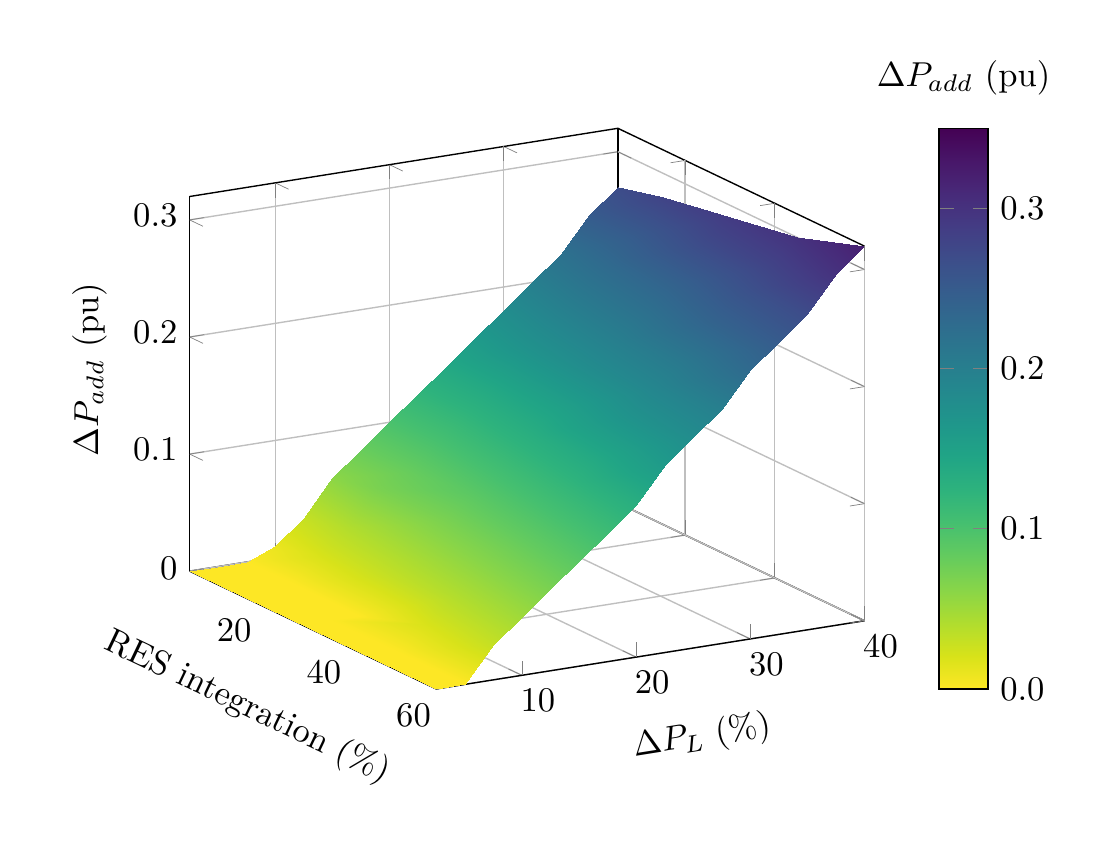}}%
	\hfil
	\subfloat[$H_{RES}$ to fulfil ROCOFs limits]{\includegraphics[height=0.21\textheight]{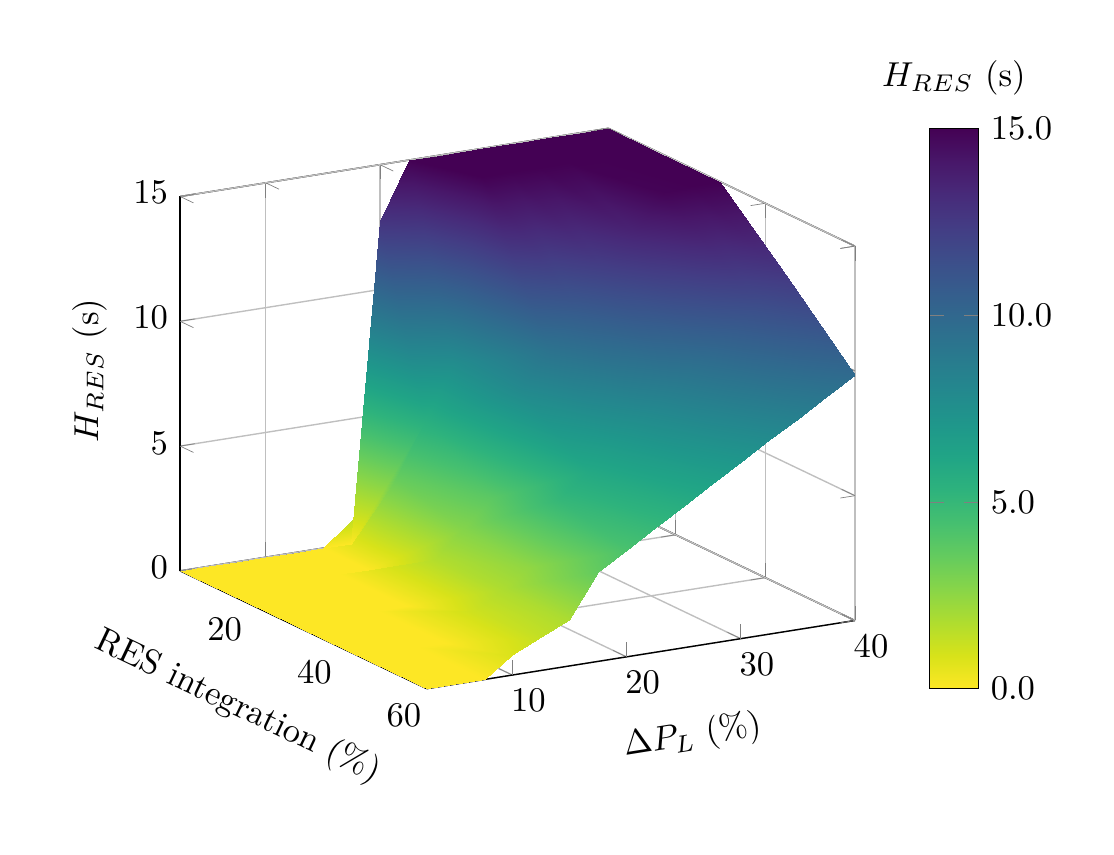}}
	\caption{\textcolor{black}{$H_{RES}$ and $\Delta P_{add}$ estimation ($H_{T}=2$~s, $H_{H}=1.75$~s). }}
	\label{fig.case3_mod}
\end{figure}

\textcolor{black}{The additional power $\Delta P_{add}$ to be provided in order to avoid the nadir frequency under $f<49.2$~Hz takes a maximum value of 0.32~pu. 
This $\Delta P_{add}$ can be given by different resources (such as including RES in frequency control, storage solutions, increasing the PFR of conventional power plants, interconnections, or a combination of some of these). As can be seen in Figures~\ref{fig.case1_mod}--\ref{fig.case3_mod}, the higher the RES integration and $\Delta P_{L}$, the higher the additional power $\Delta P_{add}$ required. In fact, a lineal regression with $R^{2}=0.96$ depending on the RES integration and the $\Delta P_{L}$ can be deduced.} 

\textcolor{black}{With regard to $H_{RES}$, some of the estimated results may lead to  unrealistic solutions. Indeed, a maximum value of $H_{RES}=96$~s was obtained. This $H_{RES}$ could also come from different resources. 
Moreover, the RES virtual inertia constant  does not have any physical meaning (and an arbitrary value could be set). According to \cite{zhang12,krpan2017inertial}, the most
appropriate value for VSWT is $1.85\cdot H_{WT}$ (being $H_{WT}$ the inertia constant of the wind turbine) to avoid the stalling of the turbine. As a result, a maximum $H_{RES}$ of 15~s is considered in Figures~\ref{fig.case1_mod}--\ref{fig.case3_mod}, which is slightly over inertia constants of conventional power plants (2--10 s)~\cite{fernandez19,fernandez19analysis}. By considering this maximum limit, ROCOF is within the limits of the ENTSO-E recommendations for 85\% of the analysed cases (612 scenarios).}

\begin{figure*}[tbp]
	\centering
	\subfloat[Thermal and hydro inertia constant scenarios for 0<$\Delta P_{L}$<20.]{\includegraphics[height=0.20\textheight]{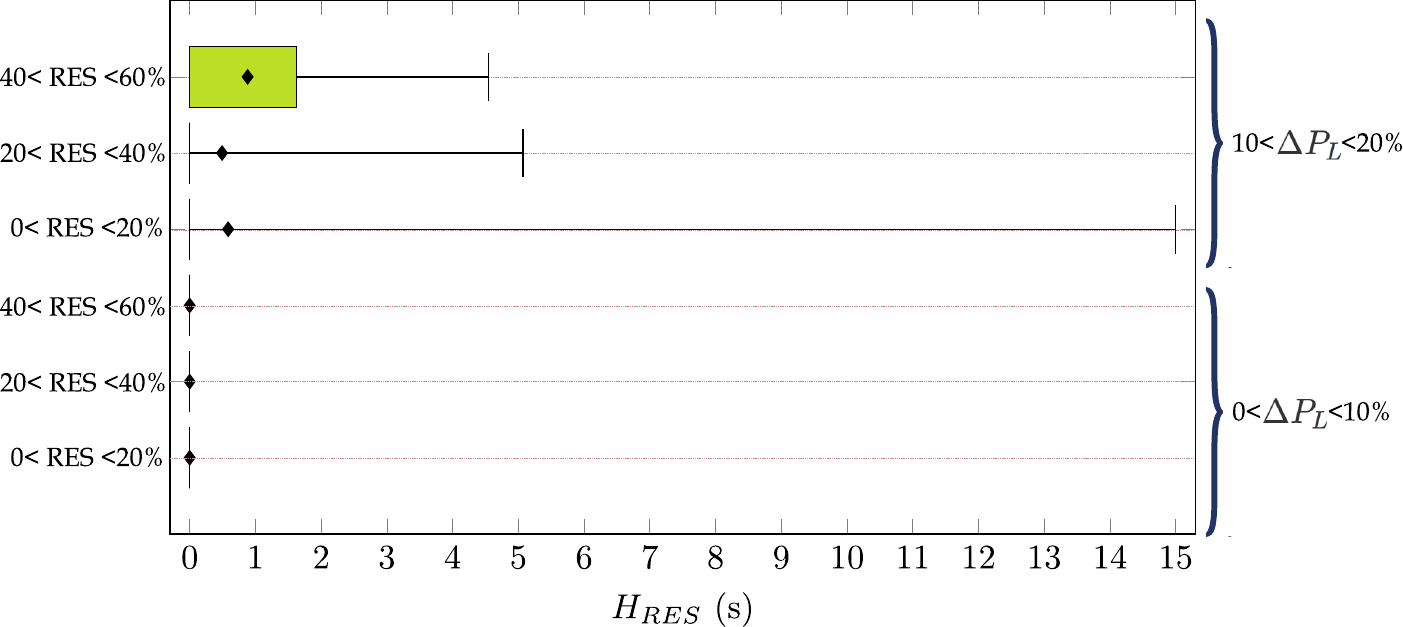}}%
	\hfil
	\subfloat[Thermal and hydro inertia constant scenarios for 20<$\Delta P_{L}$<40.]{\includegraphics[height=0.20\textheight]{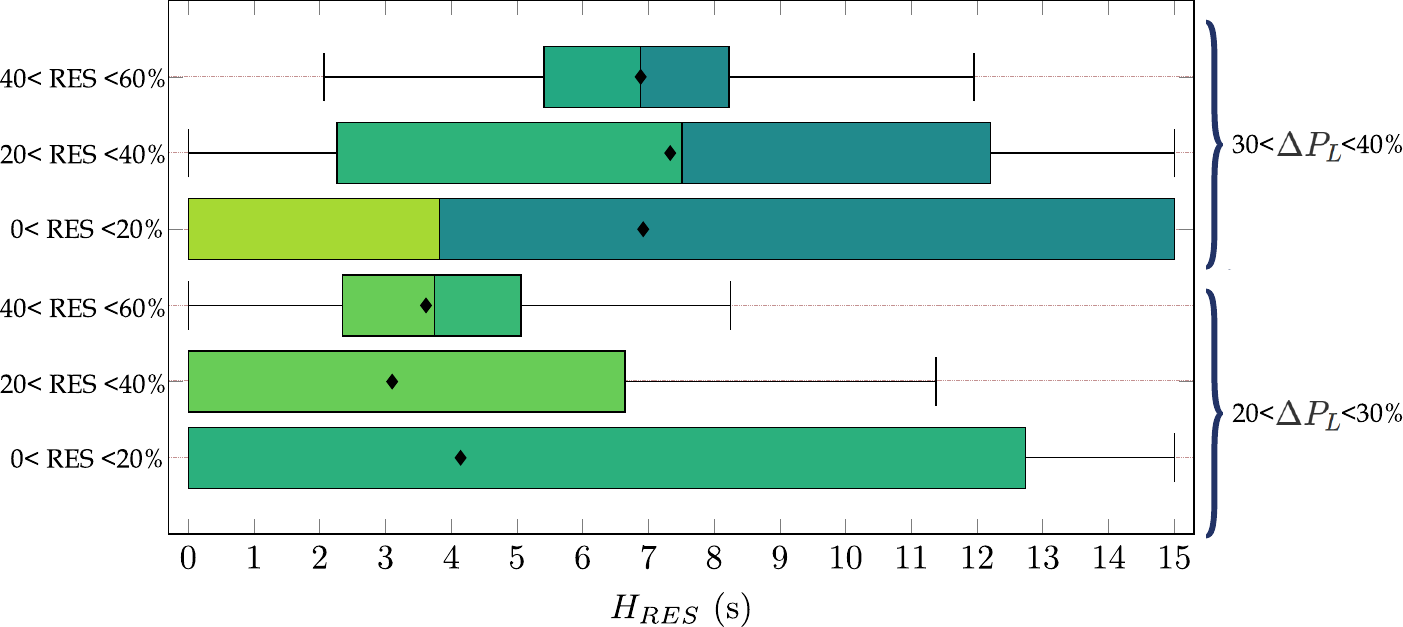}}
	\caption{\textcolor{black}{$H_{RES}$ to fulfil ROCOFs limits: summary of virtual inertia values.}}
	\label{fig.HRES_cajas_bigotes}
\end{figure*}

\begin{figure*}[tbp]
	\centering
	\subfloat[Thermal and hydro inertia constant scenarios for 0<$\Delta P_{L}$<20.]{\includegraphics[height=0.20\textheight]{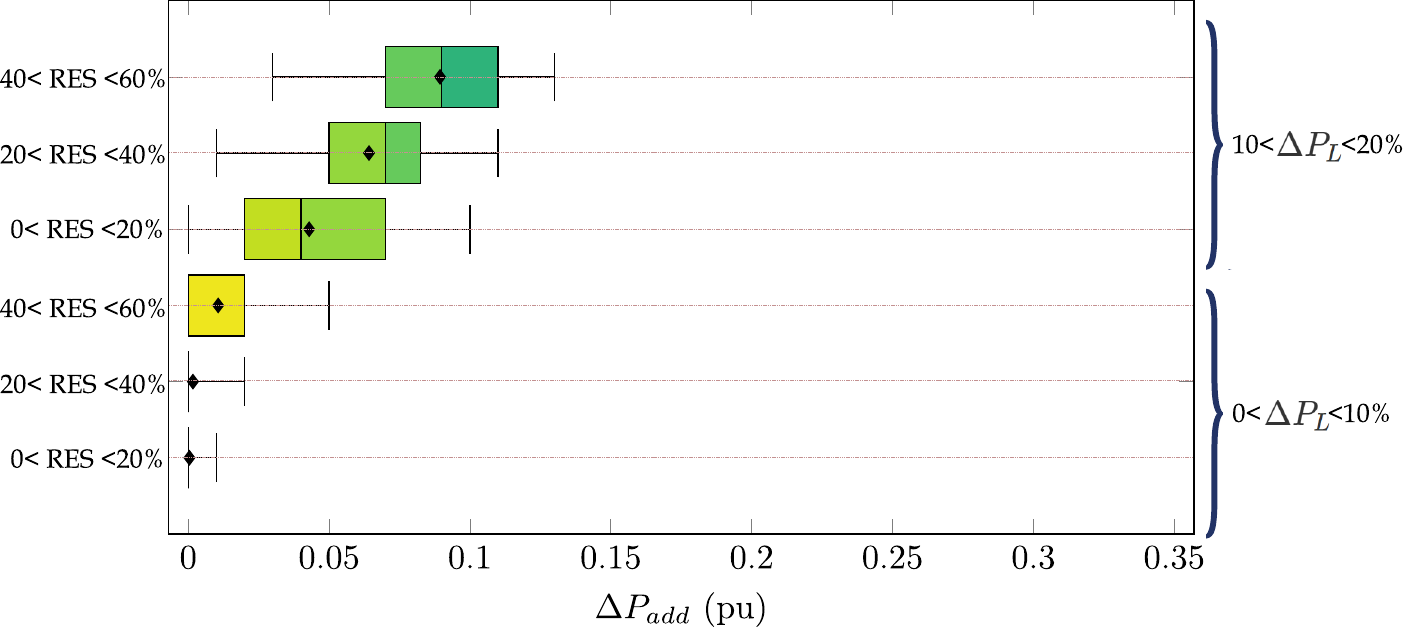}}%
	\hfil
	\subfloat[Thermal and hydro inertia constant scenarios for 20<$\Delta P_{L}$<40.]{\includegraphics[height=0.20\textheight]{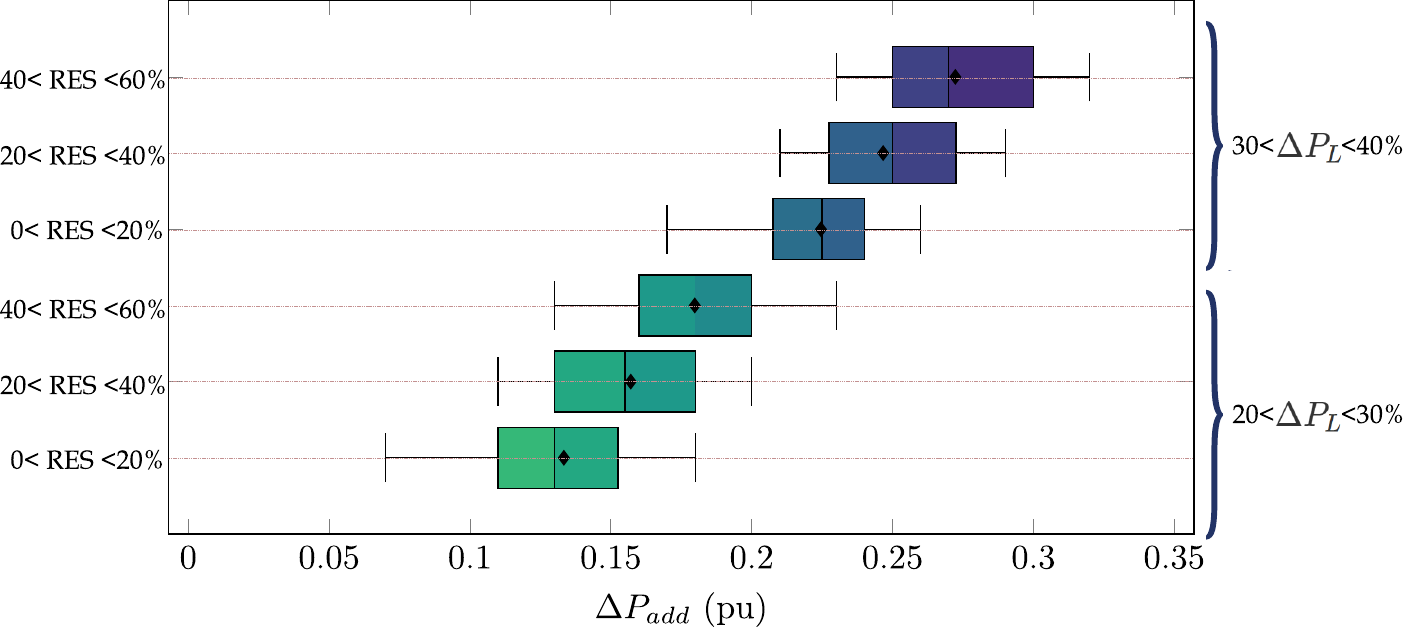}}
	\caption{\textcolor{black}{$\Delta P_{add}$ to fulfil nadir frequency limit: summary of values.}}
	\label{fig.Padd_cajas_bigotes}
\end{figure*}

\subsection{Discussion}
\label{sec.evaluation_results}

\textcolor{black}{
The recommended inertia constant of future power systems, according to ENTSO-E, ranges in between 2--3 s \cite{entsoe_high}, whereas in conventional power plants it used to be 5--6 s \cite{entsoe_si}. In this study, authors have considered cases in which $H_{eq}$ is lower and higher than such recommended values, corresponding to a wide range of RES. In fact, under the same RES integration percentage, a relevant variety of $H_{eq}$ is obtained, highly depending on the $H_{T}$ and $H_{H}$, refer to Table \ref{tab.H_eq}. As a result, the different values of $H$ for conventional power plants is identified as a major importance to properly carry out frequency deviation studies for future scenarios.}

\textcolor{black}{As was mentioned in Section {\ref{sec.introduction}}, PV and wind resources are considered as the most promising RES technologies. The absence of rotational parts in PV power plants implies an inertia constant $H\approx0$ \cite{tielens17phd}. The stored kinetic energy and, subsequently, deployable inertia of VSWTs is not directly available due to the power inverter \cite{gonzalez2014frequency}. As a consequence, the more the RES penetration into a power system, the weaker it will be in terms of frequency stability due to the reduction of $H_{eq}$ \cite{pulgar2017inertia,fang2018improved}. Consequently, the maximum penetration of RES for this $H_{eq,min}$ of 2--3 s is between 50-70\%. In fact, some authors affirm that with current RES technologies, only 50\% of the overall electricity demand can be provided by them~\cite{zakeri15,weitemeyer15}. This RES integration discussion should also include an extensive frequency response analysis, which is the aim of this work. The 720 scenarios previously analysed have been then simulated by including the results of $H_{RES}$ and $\Delta P_{add}$ estimated in Section \ref{sec.including_hres_padd}. In this way, it is then possible to verify that $H_{RES}$ and $\Delta P_{add}$ values are suitable to fulfil the ENTSO-E recommendations. Figure {\ref{fig.HRES_cajas_bigotes}} summarises in a box-and-whisker plot the virtual inertia values for all thermal and hydro inertia constant scenarios depending on the RES integration and the power imbalance. A maximum 15 s for virtual inertia is assumed by the authors to achieve such requirements. Additionally, Figure {\ref{fig.Padd_cajas_bigotes}} depicts the corresponding $\Delta P_{add}$ values in a box-and-whisker plot.}

{\textcolor{black}{Finally, and as examples of these frequency excursion analysis, some scenarios are depicted in Figures \ref{fig.evaluation1}--\ref{fig.evaluation2}, including the frequency evolution and the three ROCOF values considering in the ENTSO-E requirements. Different inertia values for $H_{T}$ and $H_{H}$, together with several imbalances $\Delta P_{L}$ and RES integration levels are shown. As can be seen, all nadirs are $f\geq 49.2$ Hz, thus staying in the dynamic accepted range of frequency variations. Moreover, the three ROCOF values are lower than their recommended limit values. As a consequence, the proposed methodology is suitable to estimate $\Delta P_{add}$ and $H_{RES}$ 
values, and it can be apply to other international power system requirements.}} 

\begin{figure}[htbp]
    \centering
    \subfloat[Frequency evolution for $\Delta P_{L}=20$\%]{\includegraphics[height=0.21\textheight]{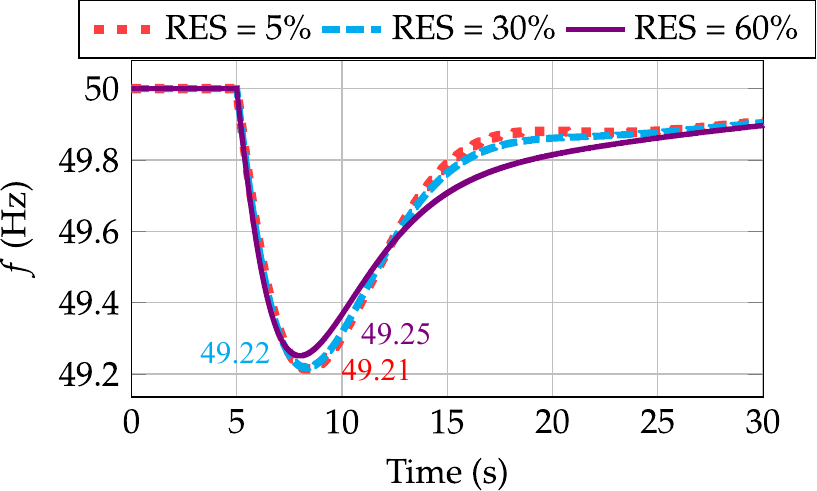}}\\
    \subfloat[ROCOF values for $\Delta P_{L}=20$\%]{\includegraphics[height=0.21\textheight]{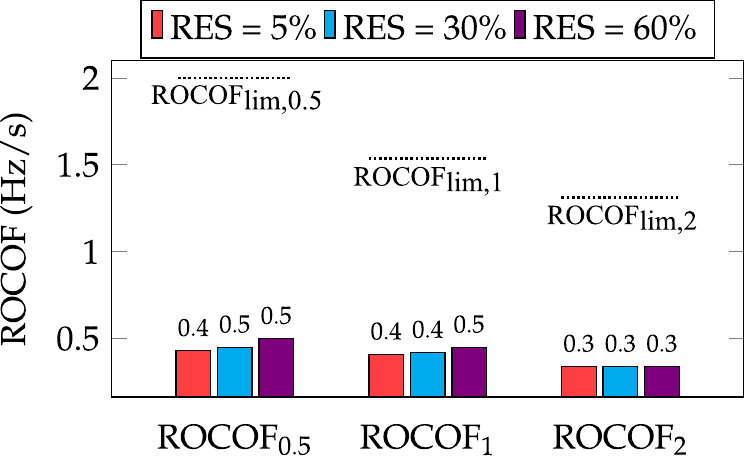}}\\
    \subfloat[$\Delta P_{L}=40$\%]{\includegraphics[height=0.21\textheight]{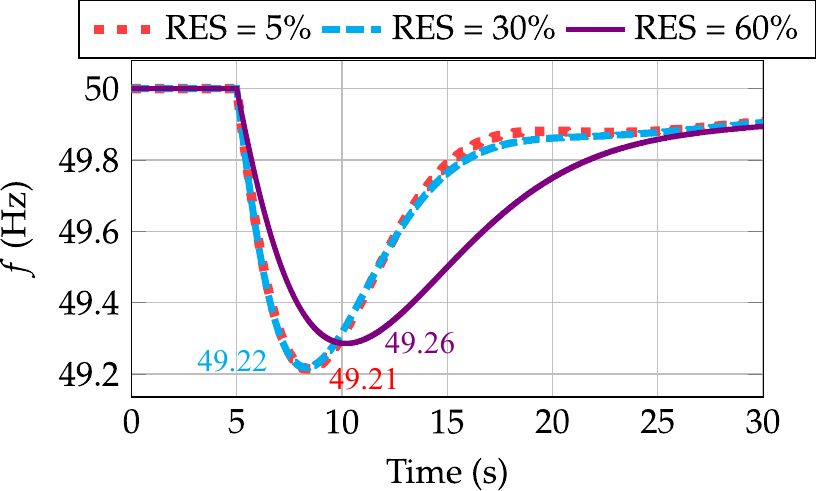}}\\
    \subfloat[ROCOF values for $\Delta P_{L}=40$\%]{\includegraphics[height=0.21\textheight]{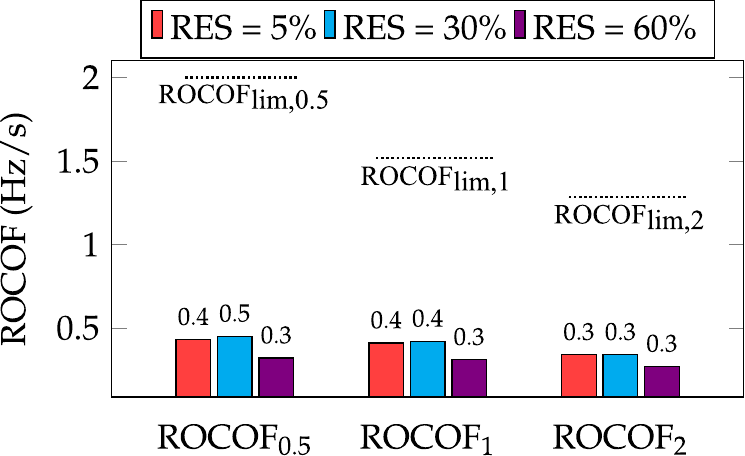}}
    \caption{Frequency evolution and ROCOF values ($H_{T}=10$ s, $H_{H}=4.75$ s).}
    \label{fig.evaluation1}
\end{figure}

\begin{figure}[htbp]
    \centering
    \subfloat[Frequency evolution for $\Delta P_{L}=12.5$\%]{\includegraphics[height=0.21\textheight]{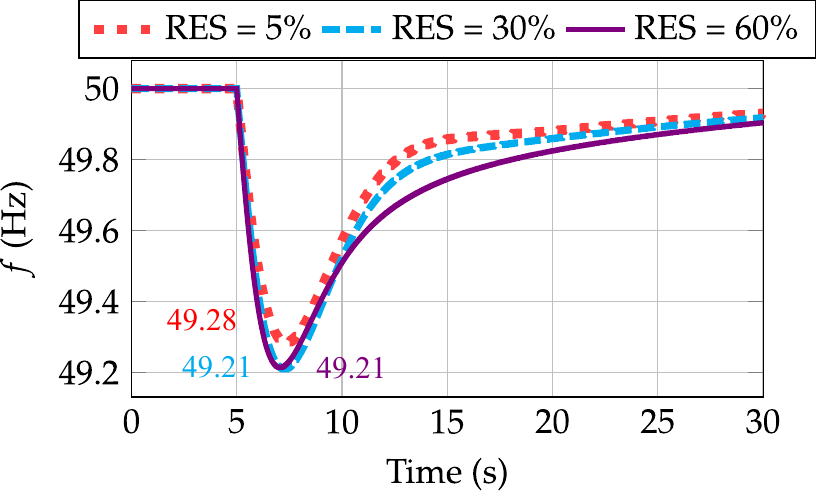}}\hfill
    \subfloat[ROCOF values for $\Delta P_{L}=12.5$\%]{\includegraphics[height=0.21\textheight]{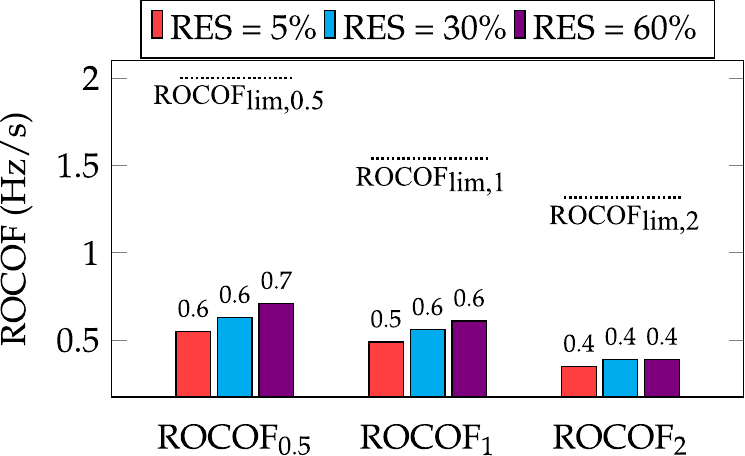}}\\
    \subfloat[$\Delta P_{L}=35$\%]{\includegraphics[height=0.21\textheight]{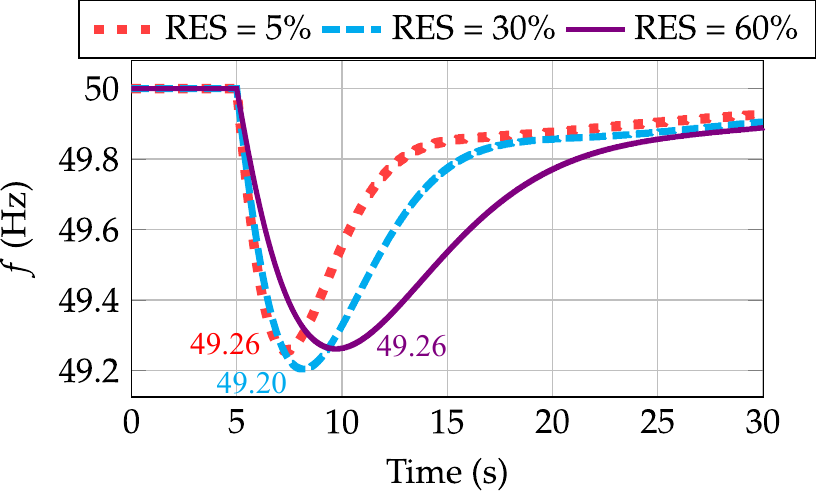}}\hfill
    \subfloat[ROCOF values for $\Delta P_{L}=35$\%]{\includegraphics[height=0.21\textheight]{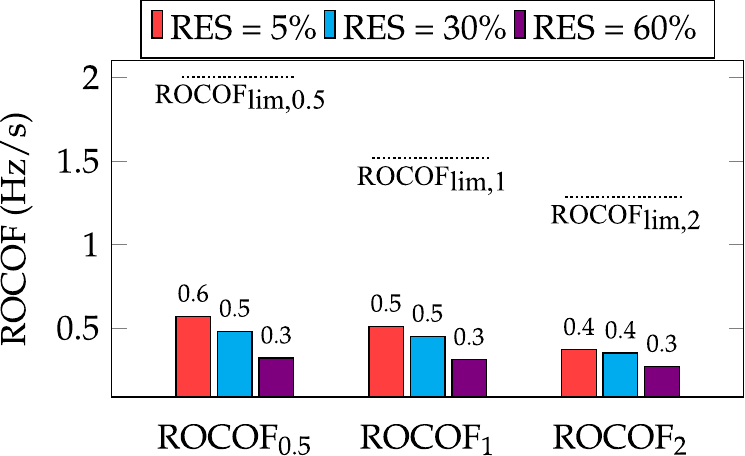}}
    \caption{Frequency evolution and ROCOF values ($H_{T}=6$ s, $H_{H}=3.25$ s).}
    \label{fig.evaluation2}
\end{figure}

\section{Conclusion}\label{sec.conclusion}

\textcolor{black}{
This paper presents a detailed analysis of nadir and ROCOFs values following ENTSO-E recommendations for the interconnected European power system. A total of 720 generation mix scenarios are simulated, considering different RES integration (between 5\% and 60\%), power imbalance and inertia constants of conventional power plants (thermal and hydro-power). 
The minimum additional power and virtual inertia required to fulfil the nadir and ROCOF limits established by ENTSO-E are estimated for each scenario. The results show that an additional power between 0 and 0.32~pu is enough to avoid any frequency excursion lower than  49.2~Hz, even if RES generation is over 50\% and the power imbalance is 40\%. This additional power can be provided by a sort of sources, such as RES (through frequency control techniques), increasing the primary frequency reserves of conventional power plants, storage systems, provided by a different power system through interconnections, or a combination of some of them. 
Authors assume a maximum virtual inertia lower than 15 s. Under this assumption, the ROCOF values following the ENTSO-E requirements are fulfilled for 85\% of the simulated scenarios. From our analysis, it can be deduced that the additional power and virtual inertia to support frequency deviations after power imbalances can be provided independently and from different resources. This is due to the fact that ROCOF is directly related to inertia, but nadir is not so affected by such parameter. This analysis is carried out considering the interconnected European power system. Nevertheless, it can be applied to any other power system with high penetration of renewable sources, considering the specific recommendations of such grid. }

\section{Acknowledgements}\label{achnowledgment}

This work is supported by the Spanish Ministry of Education, Culture and Sport ---FPU16/04282--- and by the Spanish Ministry of Economy and Competitiveness and the European Union ---FEDER Funds, ENE2016-78214-C2-1-R---.

\bibliographystyle{iet}
\bibliography{biblio2}


\end{document}